\documentclass[a4paper,11pt]{article}
\renewcommand{\baselinestretch}{1} 
\usepackage{setspace}
\usepackage{amsmath}
\usepackage{graphics,epsfig}
\usepackage{palatino}
\usepackage{amstext,amsfonts,amssymb,amsthm,amsmath,mathrsfs,epic,eepic}
\usepackage{subfigure}  
\usepackage{authblk}
\usepackage{lineno}
\usepackage[hang,flushmargin]{footmisc} 
\usepackage{multicol}

\usepackage{float}
\usepackage{mathrsfs}
\usepackage{microtype}
\usepackage[english]{babel}
\usepackage{booktabs}
\usepackage{chngcntr}
\usepackage{mathtools}
\usepackage{lineno}

\usepackage{bm}
\usepackage{geometry, amsmath}
\usepackage{xcolor}
\definecolor{TangoButter1}{HTML}{FCE94F}
\definecolor{TangoButter2}{HTML}{EDD400}
\definecolor{TangoButter3}{HTML}{C4A000}
\definecolor{TangoOrange1}{HTML}{FCAF3E}
\definecolor{TangoOrange2}{HTML}{F57900}
\definecolor{TangoOrange3}{HTML}{CE5C00}
\definecolor{TangoChocolate1}{HTML}{E9B96E}
\definecolor{TangoChocolate2}{HTML}{C17D11}
\definecolor{TangoChocolate3}{HTML}{8F5902}
\definecolor{TangoChameleon1}{HTML}{8AE234}
\definecolor{TangoChameleon2}{HTML}{73D216}
\definecolor{TangoChameleon3}{HTML}{4E9A06}
\definecolor{TangoSkyBlue1}{HTML}{729FCF}
\definecolor{TangoSkyBlue2}{HTML}{3465A4}
\definecolor{TangoSkyBlue3}{HTML}{204A87}
\definecolor{TangoPlum1}{HTML}{AD7FA8}
\definecolor{TangoPlum2}{HTML}{75507B}
\definecolor{TangoPlum3}{HTML}{5C3566}
\definecolor{TangoScarletRed1}{HTML}{EF2929}
\definecolor{TangoScarletRed2}{HTML}{CC0000}
\definecolor{TangoScarletRed3}{HTML}{A40000}
\definecolor{TangoAluminium1}{HTML}{EEEEEC}
\definecolor{TangoAluminium2}{HTML}{D3D7CF}
\definecolor{TangoAluminium3}{HTML}{BABDB6}
\definecolor{TangoAluminium4}{HTML}{888A85}
\definecolor{TangoAluminium5}{HTML}{555753}
\definecolor{TangoAluminium6}{HTML}{2E3436}

\begin{document}
\newcommand{\bra}{\langle}
\newcommand{\ket}{\rangle}
\newcommand{\be}{\begin{eqnarray}}
\newcommand{\ee}{\end{eqnarray}}
\newcommand{\del}{\partial}
\newcommand{\E}{\mbox{erf}}
\renewcommand{\eqref}[1]{Eq.~(\ref{#1})}
\newcommand{\eqsref}[1]{Eqs.~(\ref{#1})}
\newcommand{\eqssref}[2]{Eqs.~(\ref{#1})-(\ref{#2})}
\newcommand{\eqnumref}[1]{(\ref{#1})}
\newcommand{\figref}[1]{Fig.~\ref{#1}}
\newcommand{\figsref}[1]{Figs.~\ref{#1}}
\newcommand{\figssref}[2]{Figs.~\ref{#1}-\ref{#2}}
\newcommand{\rrule}[1]{\rule[#1]{0pt}{0pt}}
\newcommand{\etal}{\emph{et al}}
\renewcommand{\baselinestretch}{1.5}

\title{\vspace{-2 cm} Exploratory Adaptation in Large Random Networks}
\author[1,2]{\large Hallel I. Schreier}

\author[3]{\large Yoav Soen}
\author[1,4,*]{\large Naama Brenner}
\date{\vspace{-5ex}}
\affil[1]{\small Network Biology Research Labratories,  Technion - Israel Institute of Technology, Haifa 32000, Israel}
\affil[2]{\small Interdisciplinary Program of Applied Mathematics,  Technion - Israel Institute of Technology, Haifa 32000, Israel}

\affil[3]{\small Department of Biological Chemistry, Weizmann Institute of Science, Rehovot 76100, Israel}
\affil[4]{\small Department of Chemical Engineering, Technion - Israel Institute of Technology, Haifa 32000, Israel} 
\affil[*]{\small Corresponding Author: nbrenner@tx.technion.ac.il}

\maketitle
\normalsize

\begin{changemargin}{1.5cm}{1.5cm} 
\setstretch{1.15}
{\bf The capacity of cells and organisms to respond to challenging conditions in a repeatable manner is limited by a finite repertoire of pre-evolved adaptive responses. Beyond this capacity, cells can use exploratory dynamics to cope with a much broader array of conditions. However, the process of adaptation by exploratory dynamics within the lifetime of a cell is not well understood. Here we demonstrate the feasibility of exploratory adaptation in a high-dimensional network model of gene regulation. Exploration is initiated by failure to comply with a constraint and is implemented by random sampling of network configurations. It ceases if and when the network reaches a stable state satisfying the constraint. We find that successful convergence (adaptation) in high dimensions requires outgoing network hubs and is enhanced by their auto-regulation. The ability of these empirically-validated features of gene regulatory networks to support exploratory adaptation without fine-tuning, makes it plausible for biological implementation.}

\end{changemargin}

\vspace{0.5cm}


\setstretch{1.2}
\begin{multicols}{2}
{\fontsize{14}{14} {\noindent \bf Introduction}}

\noindent The ability to organize a large number of interacting processes into persistently viable states in a dynamic environment is a striking property of cells and organisms. Many frequently encountered perturbations  (temperature, osmotic pressure, starvation and more), trigger reproducible adaptive responses \cite{Gasch2000,Causton2001,LopezMaury2008}. These were assimilated into the organism by variation and selection over evolutionary time. Despite the large number and flexible nature of these responses, they span a finite repertoire of actions and cannot address all possible scenarios of novel conditions. Indeed, cells may encounter severe, unforeseen situations within their lifetime, for which no effective response is available. To survive such challenges, a different type of ad-hoc response can be employed, utilizing exploratory dynamics \cite{KirschnerGerhart,Braun2015,Soen2015}. 

The capacity to withstand unforeseen conditions was recently demonstrated and studied using dedicated experimental models of novel challenge in yeast
\cite{Stern2007,David2010,Katzir2012} and flies \cite{Stern2012}. Adaptive responses exposed in these experiments involved transient changes in the expression of hundreds of genes,  followed by convergence to altered patterns of expression. Analysis of repeated experiments showed that a large fraction of the transcriptional response can vary substantially across replicate trajectories of adaptation \cite{Stern2007,Katzir2012}. These findings suggest that coping with unforeseen challenges within one or a few generations relies on induction of exploratory changes in gene regulation over  time in an individual 
\cite{Braun2015,Soen2015}. 

Several properties of gene regulatory networks may support such exploratory adaptation. These include a large number of potential interactions between genes \cite{Tong2004}, context-dependent plasticity of interactions 
\cite{Harbison2004,Luscombe2004,Niklas2015,Bondos2016flexibility} and multiplicity of microscopic configurations consistent with a given phenotype  \cite{Weiss2000}. Despite these properties, the feasibility of acquiring adaptive phenotypes by random exploration within a single organism remains speculative and poorly understood. In particular, it is not known how exploration may converge rapidly enough in the high dimensional space of possible configurations?  what determines the efficiency of this exploration? and what ensures the stabilization of new phenotpes? 

Here we address these open questions by introducing a network model of gene regulation, which demonstrates the capacity to adapt by exploratory dynamics in a single cell (as opposed to selection on existing variation in a population). Exploration is triggered by failure to satisfy a newly-imposed external demand, and is implemented by a random walk in the space of network configurations. Exploration relaxes if and when the system reaches a stable state satisfying this demand. We show that the success of this exploratory adaptation in high dimension requires that the network include outgoing hubs. Adaptive capability is further enhanced by autregulation of these outgoing hubs. Since these are both well-known properties of gene regulatory networks, our findings establish a basis for a biologically plausible mode of adaptation by exploratory dynamics. 


\bigskip
{\fontsize{14}{14} {\noindent \bf Results}}\\
\noindent \textbf{\textit{Exploratory Adaptation Model}}\\
\noindent To investigate the feasibility of exploratory adaptation, we introduce a model of gene regulatory dynamics incorporating random changes over time in a single network.
The model consists of a large number, $N$, of microscopic components $\mathbf{{x}}\!=\!(x_1,x_2...x_N)$, governed by the following nonlinear equation of motion (Fig. 1A):
\be
\label{eq:Dynamics}
\mathbf{\dot{x}} &=&W\phi(\mathbf {x})-\mathbf {x},  
\ee
\noindent where $W$ is a random matrix, representing the intracellular network of interactions; $\phi(\mathbf {x})$ an element-wise saturating function restricting the dynamic range of the variables; and the relaxation rates are set to unity. Previous work has used similar equations to address
evolutionary aspects of gene regulation \cite{Kauffman1993,Wagner2011} as well as interactions and relaxation in neuronal networks \cite{Amit92}. Most studies have focused on networks with uniform (full or sparse) connectivity; much less is known about the dynamics for networks with 
non-uniform topological structures, which may be of relevance to gene regulation. 

Here we consider sparse random networks with different types of topological properties.
For all cases, the interaction matrix $W$ is composed of an element-wise (Hadamard) product, 
\be
W = T\circ J,
\ee
\noindent where $T$ is a random topological backbone (adjacency) matrix with binary (0/1) entries representing potential interactions between network elements; and J is a random matrix specifying the actual interaction strengths. To represent context-dependent regulatory plasticity, we assume that the backbone remains fixed, whereas the interaction strengths are plastic and amenable to change over time. We will emphasize below network sizes and topological structures that are relevant to gene regulatory networks.

On a macroscopic level, we consdier a cellular phenotype, $y$, which depends on the microscopic components and can affect the cell's functionality and state of stress. We define this phenotype as a linear combination of microscopic variables 
\be
y(t)=\mathbf {b}\!\cdot\! \mathbf {x}(t)
\ee
\noindent  with an arbitrary vector of coefficients $\mathbf b$. To model an unforeseen challenge, the system is subjected to an arbitrary contstraint of maintaining the phenotype in a given range $y(t)\!\approx\! y^*$. Importantly, any given value of the phenotype can be realized by a large number of alternative microscopic combinations. 

Deviation from compliance with the constraint is represented by a
global cellular function ${\cal M}(y\!-\!y^*)$, corresponding to the level of mismatch between the current phenotype and the demand. 
This mismatch is effectively zero inside a "comfort zone" of size $\varepsilon$ around $y^*$ and increases sharply beyond it. Biologically, the comfort zone can be interperted as a range of phenotypes that can be tolerated in a given environment without invoking significant stress. This is represented mathematically by a range of values which satisfy the constraint 
(in contrast to many optimization problems which require adherence to a specific value). 

When the phenotype deviates from the comfort zone, the mismatch drives an exploratory search 
realized by small random changes in the interaction strengths, forming a random walk in the elements of the matrix $J$:

\begin{align}
 dJ_t = \sqrt{D \cdot {\cal M}(y\!-\!y^*)} \cdot d{\cal W}_t . \quad J(t\!=\!0)\!=\!J_0, 
\end{align}

\noindent  where ${\cal W}_t$ is the standard Wiener process. The amplitude of the random walk is controlled by a scale parameter, $D$, and the mismatch level, ${\cal M}$. These random changes can arise from diverse sources of variation affecting the levels of transcription regulators \cite{LopezMaury2008,Furusawa2008generic,Kaneko2013}, as well as their regulatory interactions (e.g. alternative splicing, conformations of transcription factors and their post-translational modifications \cite{Niklas2015,Bondos2016flexibility}).

The random walk constitutes an exploratory search for network configurations in which the dynamical system  in Eq. (1) satisfies the constraint in a stable manner. Random occurrence of such a configuration decreases the search amplitude, thereby promoting  its relaxation by reducing the drive for exploration \cite{Soen2015,ShahafMarom2001}. Convergence of this process to a stable state satisfying the constraint is not {\it a-priori} guaranteed. Intuitively, it may be expected that  randomly varying a large number of parameters in a nonlinear high-dimensional system will cause the dynamics to diverge. Surprisingly, we find that the adaptation process can in fact converge; however, as shown below, convergence depends on key topological properties of the network.

\bigskip
\noindent \textbf{\textit{Adaptation Depends on Network Topology}}

\noindent An example of adaptive convergence is shown in Fig. 1B-D. At $t=0$, the system is confronted with a demand and starts an exploratory process in which the connection strengths are slowly modified. Fig. 1B displays the time trajectories of four of these connection strengths. During this exploration, the microscopic variables, $\bm{\mathbf x}$, and the phenotype, $y$, exhibit highly irregular behavior, rapidly sampling a large dynamic range (Figs. 1C and 1D respectively). At $t=400$, the system manages to stably reduce the mismatch to zero and converges to a fixed point (Fig. 1) or a small-amplitude limit-cycle (Supplementary Note 3, Convergence to a limit cycle), and remain within the comfort zone $\pm  \varepsilon$ around $y^*$.  The state of convergence is found to be a stable attractor that is robust against small perturbations of the dynamic variables, $\bm {\mathbf{x}}$, and the interactions strengths, $W_{ij}$ (Supplementary Note 3, Stability of the adapted state). The differences between the amplitude of temporal changes in Figs. 1B and 1C,D reflects the separation of timescales between the slowly accumulating changes in interaction strengths, governed by the small value of $D$ in Eq. (4), and the intrinsic dynamics of Eq. (1). 

\begin{figure*}
	\begin{center}
			\includegraphics[width=16cm]{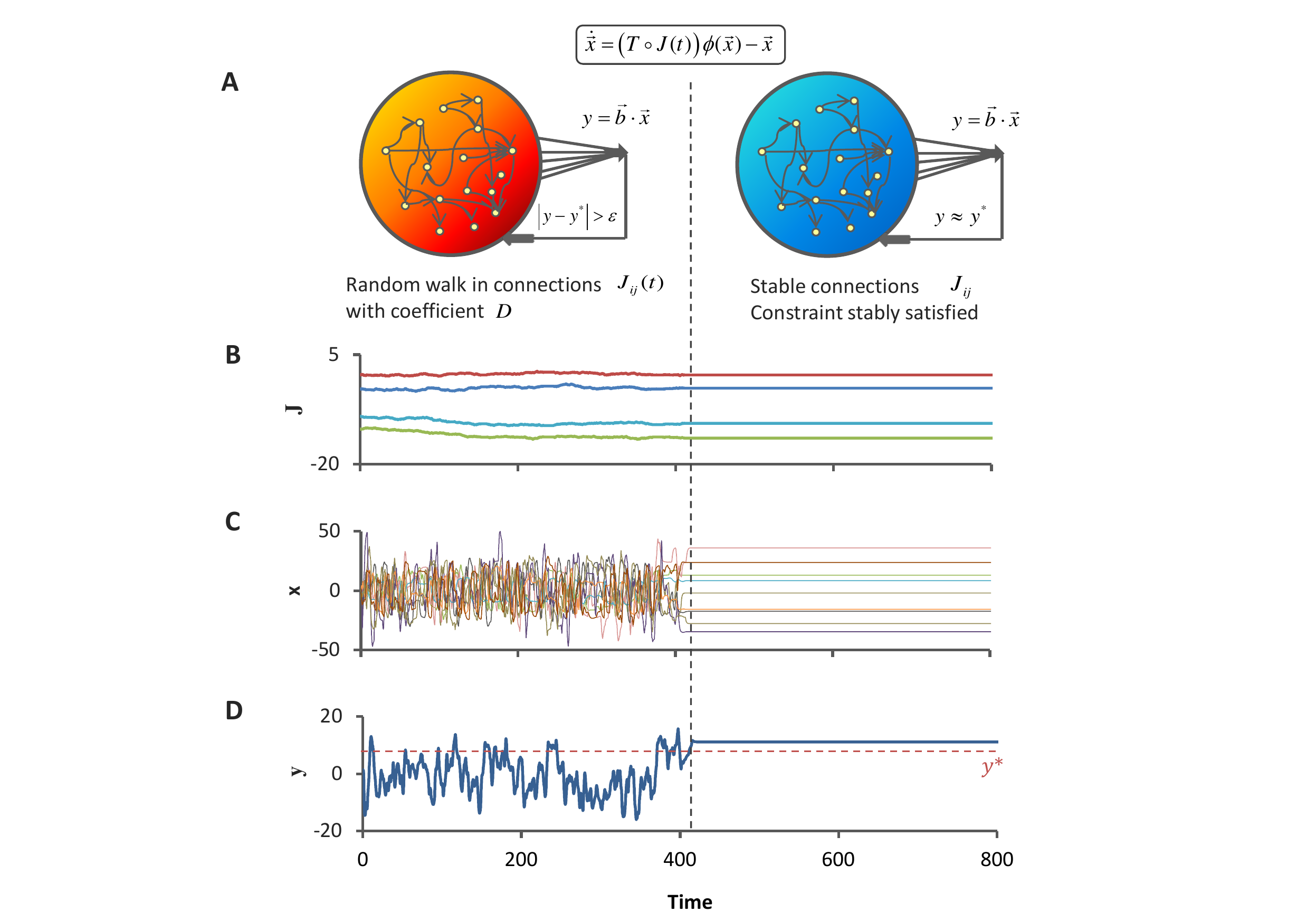}
		\caption{{\bf Exploratory dynamics and convergence to a constraint-satisfying stable state}. {\bf (A)}. Schematic representation of the model: a random $N\times N$ network, composed of an adjacency matrix $T$ and an interaction strength matrix $J$, governs a nonlinear dynamical system (equation in box; $\phi(x)=\tanh(x))$. The resulting spontaneous dynamics are typically irregular for large enough interactions. A macroscopic variable, the phenotype $y$, is subject to an arbitrary constraint $y\!\approx \!y^*$ with finite precision $\varepsilon$. When the constraint is not met (left; "hot" regime), the connections strengths $J_{ij}$ undergo a random walk with magnitude determined by the coefficient $D$ and the mismatch function ${\cal M}(y\!-\!y^*)$. The random walk stops when the mismatch is stably reduced to zero (right; "frozen" regime). {\bf (B-D)} Example of exploration and convergence. Shown are representative trajectories of connection strengths (B), microscopic variables (C) and the phenotype $y$ (D) before and after convergence to a stable state satisfying the constraint. The network in this example has scale-free (SF) out-degree distribution ($a\!=\!1$, $\gamma\!=\!2.4$) and Binomial in-degree distribution ($p\simeq \dfrac{3.5}{N}$, $N$). $N\!\!=\!\!1000$,  $y^*\!\!=\!\!10$, $D\!\!=\!\!10^{-3}$, $g_0\!\! =\!\! 10$. See Methods for more details.}
		\label{fig:Convergence}
	\end{center}
\end{figure*}

Convergence of exploratory adaptation depends crucially on the topological structure of the network. To quantify this dependence we constructed random matrix ensembles with different topological backbones, manifested by distinct in- and out-going degree distributions \cite{Newman2003}. Each ensemble was evaluated with respect to the probability of convergence, estimated as the fraction of simulations which converged within a given time window.  Fig. 2A compares ensembles of networks with in- and out-degrees drawn from Binomial (‘Binom’), Exponential (‘Exp’) and Scale-Free (‘SF’) distributions. It shows high fractions of convergence, 0.5 or higher, only for ensembles with SF out-degree distributions. In contrast, the in-degree distribution affects convergence only mildly. For example, the convergence fraction (CF) of networks with SF out-degree and Binomial in-degree distributions (dark blue) is 0.5, and only 0.03 in the transposed case (light blue). This asymmetry between outgoing and incoming connections indicates that convergence of exploratory adaptation does not rely on spectral properties of the interaction matrix ensemble.

\begin{figure*}
	\begin{center}
		\includegraphics[width=16cm]{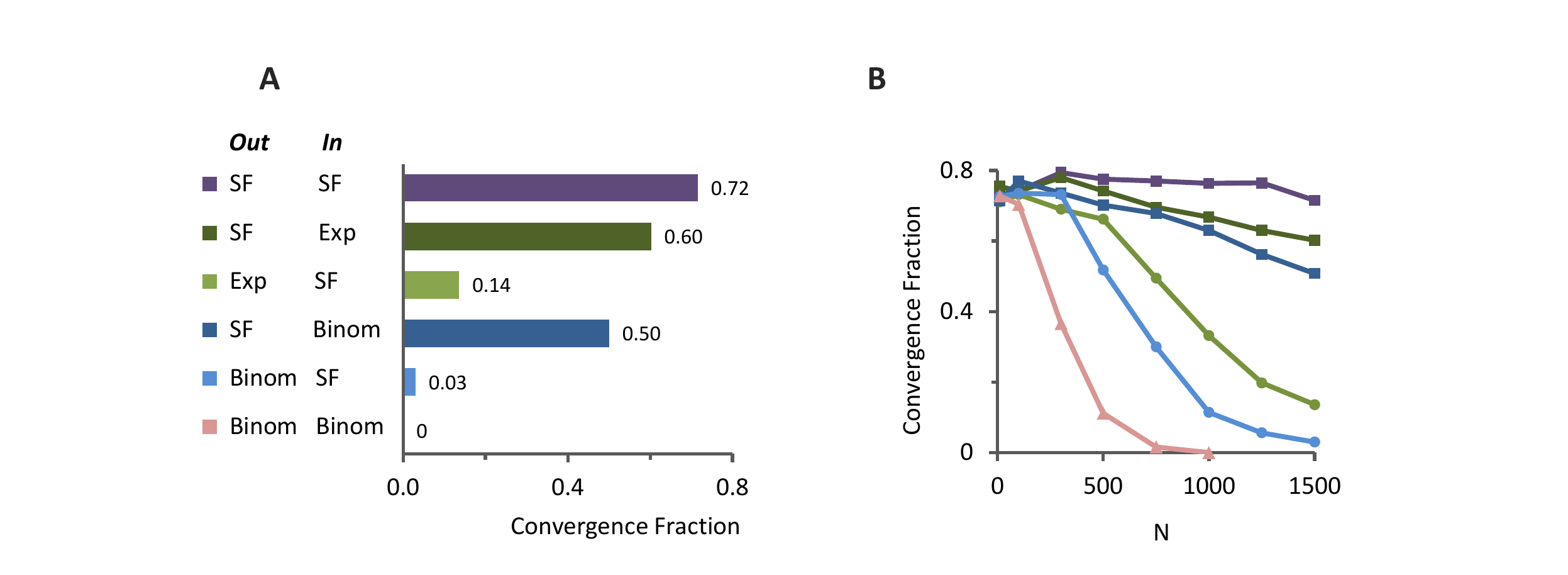}
		\caption{{\bf Convergence Fractions (CFs) depend on network topology.} {\bf (A)} Seven ensembles of networks of size $N\!\!=\!\!1500$ and different topologies exhibit remarkably different CFs. Ensembles are characterized by the out- and in- degree distributions of the adjacency matrix $T$: 'SF', scale free distribution; 'Exp', exponential distribution; 'Binom', Binomial distribution.  {\bf (B)} CF as a function of network size for the same  ensembles of (A) with matching colors. 
			$N\!=\!1500$, $y^*\!=\!0$,  $g_0 \!=\! 10$, $D\!=\!10^{-3}$. Parameters for degree distributions: SF, ($a=1$, $\gamma=2.4$); Binom, ($p\simeq \dfrac{3.5}{N}$, $N$); Exp, $\beta = 3.5$. 						 }
		
		\label{fig:Convergence}
	\end{center}
\end{figure*}

Analysis of convergence as a function of network size shows that the effect of topology becomes pronounced for large networks (Fig. 2B). The CF in small to intermediate-sized networks  ($N\!\!\lesssim\!\! 200$)  is higher and relatively independent of topology. However, as $N$ increases towards sizes that are relevant to genetic networks, the benefit of having SF out-degree distribution becomes progressively prominent. 

\bigskip
\noindent \textbf{\textit{Outgoing Hubs Enable Adaptation in Large Networks}}

Among the topological ensembles tested, an outgoing SF degree distribution was found to be crucial for convergence of large enough networks. Such distributions are characterized by a broad range of heterogeneous connectivities, with a small number of extremely highly connected nodes (hubs). To evaluate the relative contribution of outgoing hubs to convergence within this ensemble, we ranked the backbones of the connectivity matrices drawn from the SF-Binom distributions according to the out-degree of the largest hub. Fig. 3A shows that the CF increases with the connectivity of the largest outgoing hub. As a second approach to characterize hub contribution, we deleted a small number of outgoing hubs from these networks \cite{Albert2000}; this 
leads to a significant reduction in CF that is not observed upon removal of randomly chosen nodes (Fig. 3B). 

\begin{figure*}
	\begin{center}
		\includegraphics[width=16cm]{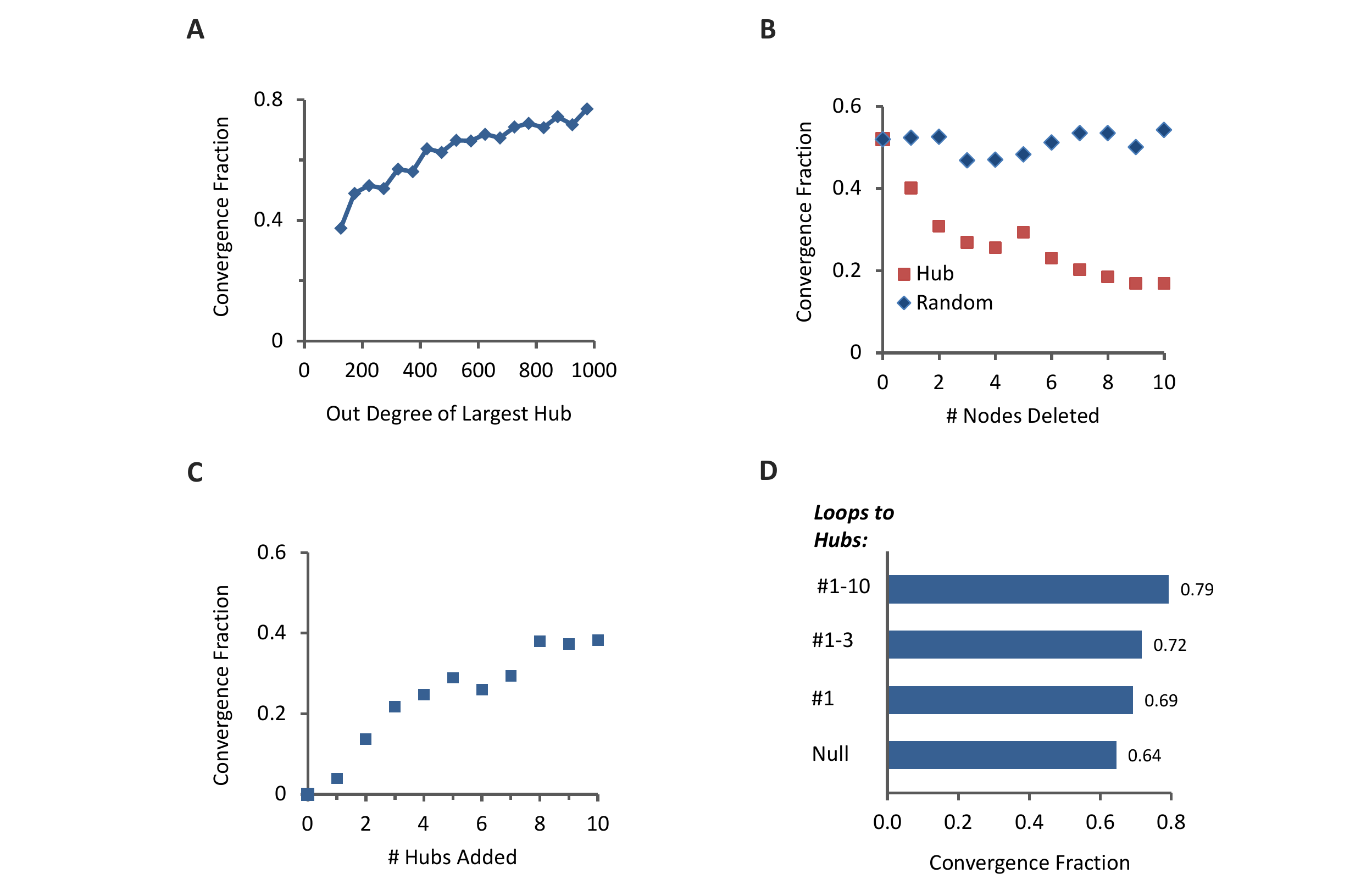}
		\caption{{\bf Exploratory adaptation depends on the existence of hubs and is enhanced by their auto-regulation.}
			{\bf (A)} CF vs. out-degree of the largest hub in a collection of SF-Binom networks binned according to their largest hub. 
			{\bf (B)} Changes in Convergence Fraction (CF) following deletion the leading hubs (red) vs. deletion of random nodes (blue) from networks with SF-Binom topology.  {\bf (C)} Effect of adding a small number of outgoing hubs to a Binon-Binom ensemble. The out-degrees of the added hubs was chosen to mimic the SF-out ensemble of Fig. 2. {\bf (D)} Effect of adding autoregulatory loops on a specific number (1,3 and 10) of the leading outgoing hubs on a background of a SF-Binom ensemble. $N\!=\!1500$, $y^*\!=\!0$,  $g_0 \!=\! 10$, $D\!=\!10^{-3}$. Parameters for degree distributions: SF, ($a=1$, $\gamma=2.4$); Binom, ($p\simeq \dfrac{3.5}{N}$, $N$); Exp, $\beta = 3.5$. 					
		} 
	\end{center}
\end{figure*}

These results indicate that, in networks of the SF-Binom ensemble, outgoing hubs have a major positive influence on the success of exploration. 
We therefore asked whether the addition of a few hubs to an otherwise poorly converging ensemble is enough to induce significant convergence. Fig. 3C indeed shows that addition of as few as 8 hubs to a Binom-Binom ensemble increases the CF from zero to about  $0.4$.

These observations are in-line with reported properties of gene regulatory networks, particularly the existence of "master regulatory" transcription factors that control the expression of hundreds of other genes \cite{Guelzim2002,Babu2004,Babu2010structure}. Since many of these master regulators are also autoregulated \cite{Pinho2014stability}, we evaluated the influence of hub autoregulation on the success of exploratory adaptation in our model. Fig. 3D shows that autoregulation of the leading hubs in the SF-Binom ensemble further increases the CFs. 

\begin{figure*}
	\begin{center}
		\includegraphics[width=16cm]{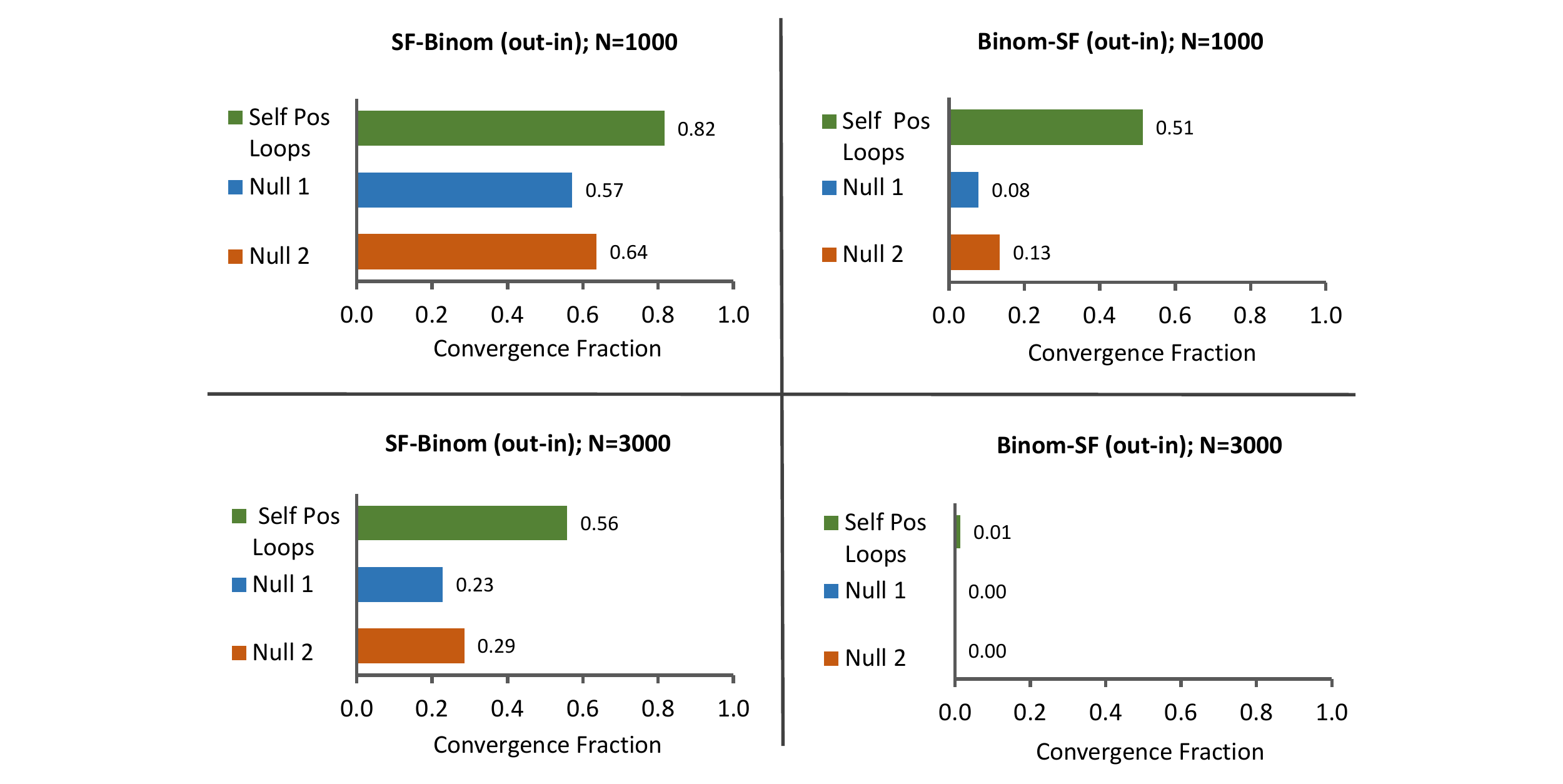}
		\caption{{\bf  Effect of positive autoregulation on convergence fractions.}
			Positive autoregulatory loops were added randomly to $10\%$ of the nodes  in four ensembles, each comprising 500 networks of a given size (N=1000 or 3000) and topology (SF-Binom or vice versa). Convergence in each ensemble is compared to controls without extra loops, with and without matching of the degree distributions to the enriched ensemble (Null 1 and Null2, respectively). Parameters of the SF and Binom distributions (prior to addition of loops) are: SF, $a=1$, $\gamma=2.4$ and Binomial, $p= \dfrac{3.5}{N}$ and $N$. Other parameters are $g_0 = 10$, $\alpha = 100$, $\mathcal M_0 = 2$, $\varepsilon = 3$, $c=0.2$, $D=10^{-3}$ and $y*=0$.  
		} 
	\end{center}
\end{figure*}

Since autoregulation motifs are commonly observed in gene regulatory networks (not only in hubs) \cite{Alon2007}, we investigated whether these motifs could also contribute to convergence when over-represented uniformly throughout the network. Fig. 4 depicts the results of adding such motifs randomly to $10\%$ of the nodes in the SF-Binom and Binom-SF ensembles. It is seen that positive autoregulation enhances convergence of for intermediate sized networks (N=1000) in both ensembles; this effect is particularly notable for the Binom-SF ensemble, which has small CF without these motifs. This contribution, however, decreases with network size and is no longer observed in the same type of networks with N=3000. We conclude that the presence of autoregulatory motifs ranodmly positioned in the network cannot substitute for hub contribution in the limit of very large networks. These results highlight the interplay of several networks properties in exploratory adaptation: network size, topology and autoregulatory motifs. The addition of common network motifs other than autoregulation did not lead to a conclusive effect on convergence (Supplementary Note 3, Dependenc of convertence on network motifs). 

\bigskip
\noindent \textbf{\textit{Adaptation Occurs Over a Wide Range of Model Parameters}}

\noindent We investigated how the capacity to adapt is affected by various model parameters. To examine the dependence on the severity of the constraint, we varied the size of the comfort zone $\varepsilon$. Fig 5A reveals a sharp decrease of the CF as $\varepsilon$ is reduced, indicating that a non-vanishing comfort zone is crucial for successful exploratory adaptation. This requirement is biologically plausible, as one expects a range of phenotypes capable of accommodating a given environment rather than a unique optimal phenotype. 
Another way of increasing the adaptation challenge is by shifting the required phenotypic range away from the origin. Reaching a shifted region is challenging because it is more rarely visited by spontaneous dynamics (Fig. 5B, grey curve). Fig. 5B indeed shows that the CF decreases as $y^*$ moves away from zero (blue curve). Importantly however, it remains much larger than the probability of encountering the required phenotype spontaneously. For example, a non-negligible convergence (CF$\sim 0.2$) is observed even for an interval around $|y^*|\!\sim \!20$ which is spontaneously encountered with probability of 0.02. 

\begin{figure*}
	\begin{center}
		\includegraphics[width=16cm]{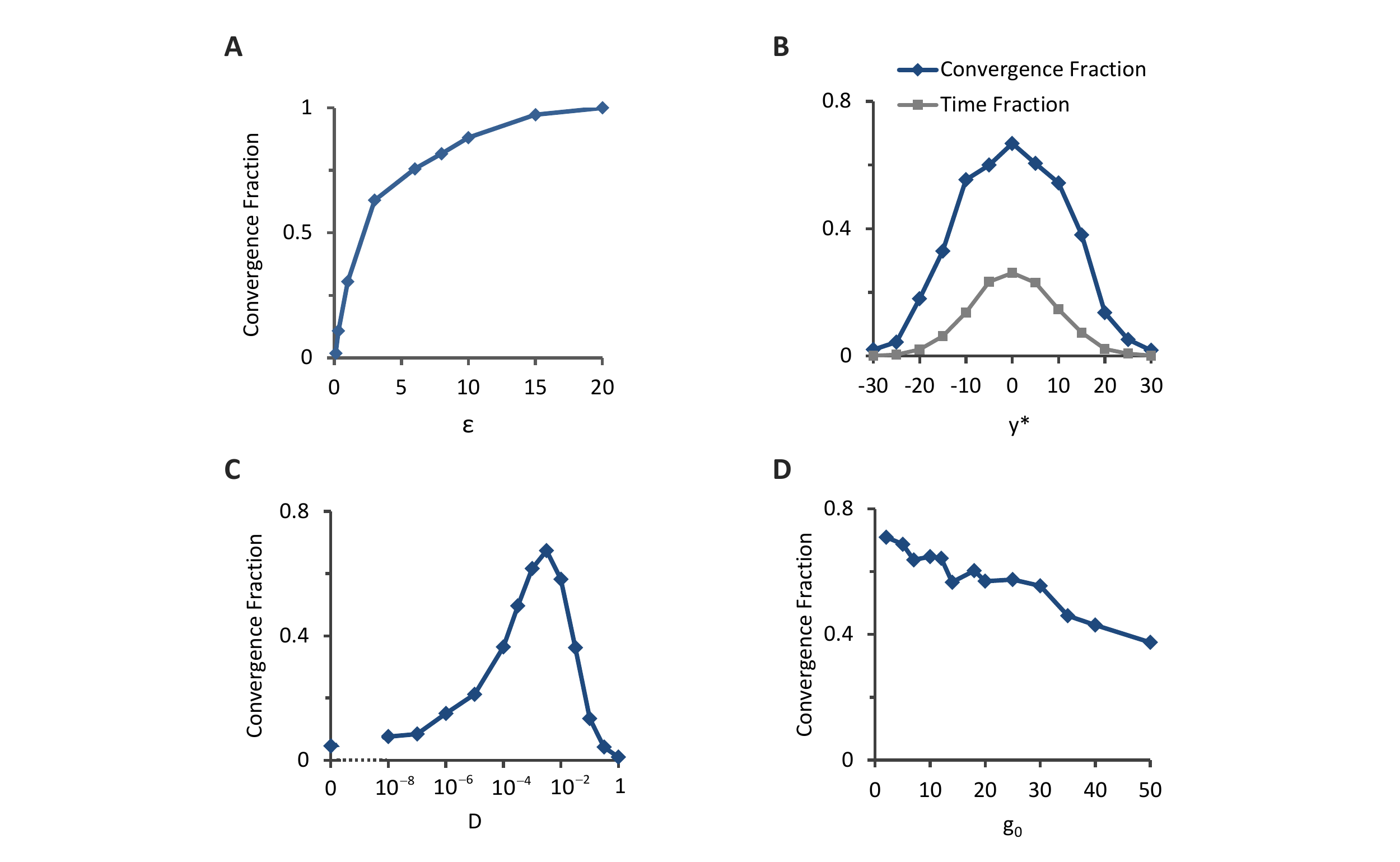}
		\caption{{\bf Dependence of CF on model parameters.}
			{\bf (A)} CF vs. $\varepsilon$, the width of the comfort zone around $y^*$. 		 
			{\bf (B)} CF (blue) vs. the constraint value $y^*$. For comparison, the grey curve shows the fraction of time in which $y(t)$ spontaneously reaches the constraint-satisfying range. 
			{\bf (C)} CF vs. the strength of exploratory random walk in connection strengths, $D$, . 
			{\bf (D)} CF vs. initial network gain (proportional to the std of connection strengths). Network ensemble with SF-out ($a=1$, $\gamma=2.4$) and Binom-in $(p\simeq \dfrac{3.5}{N}$, $N)$ degree distributions. Unless otherwise specified, all ensembles have $N\!=\!1000$,  $y*\!=\!0$,  $g_0 \!=\! 10$,  and  $D\!=\!10^{-3}$.  
		} 
	\end{center}
\end{figure*}

To evaluate the sensitivity of adaptation to exploration speed, we varied the effective diffusion coefficient in the space of connection-strengths, $D$. Fig. 5C shows that a nonzero convergence fraction is achieved for a wide range of this parameter and remains between $0.2\!-\!0.7$ over more than $5$ orders of magnitude. As the value of $D$ increases beyond a certain level where the separation of timescales ceases to hold, the convergence fraction decreases rapidly.

For a given adjacency matrix $T$, 
interactions within the network are determined by the connections strengths, $J_{ij}$. These are initially drawn from a Gaussian distribution with a zero mean and a given standard deviation. The normalized standard deviation, $g_0$ (also called network gain) determines the contribution of the first vs. second term in eq (1). 
In large homogeneous networks, this parameter has a strong effect on the dynamics of eq (1) \cite{Sompolinsky88}. In contrast, we find that the capacity to adapt by exploration in our model is relatively weakly dependent on $g_0$ (Fig. 5D).

\bigskip
\noindent \textbf{\textit{Broad, Non-Exponential Distributions of Adaptation Times}}

\noindent The analysis presented so far was based on convergence fractions within a fixed time interval. To characterize the temporal aspects of exploratory adaptation, we evaluated the distribution of convergence times in repeated simulations. Fig. 6 reveals a broad distribution ($CV\!\approx\! 1.1$), well fitted by a stretched exponential (see Supplementary Note 3, Stretched exponential fit to the distribution of convergence times). Such distributions are common in complex systems \cite{Sornette98} and were suggested to reflect a hierarchy of timescales \cite{Palmer84}. Similarly shaped distributions were found in all topological ensembles tested; however, networks with SF out-degree distributions typically converged faster than their transposed ensembles (Fig. 6A). Moreover, deletion of a small number of  leading outgoing hubs causes a significant shift towards longer convergence times (Fig. 6B). Thus, networks with larger heterogeneity in out-degrees are both more likely to converge within a given time window (Figs. 2,5), and typically converge faster (Fig. 6).

\begin{figure*}
	\begin{center}
			\includegraphics[width=16cm]{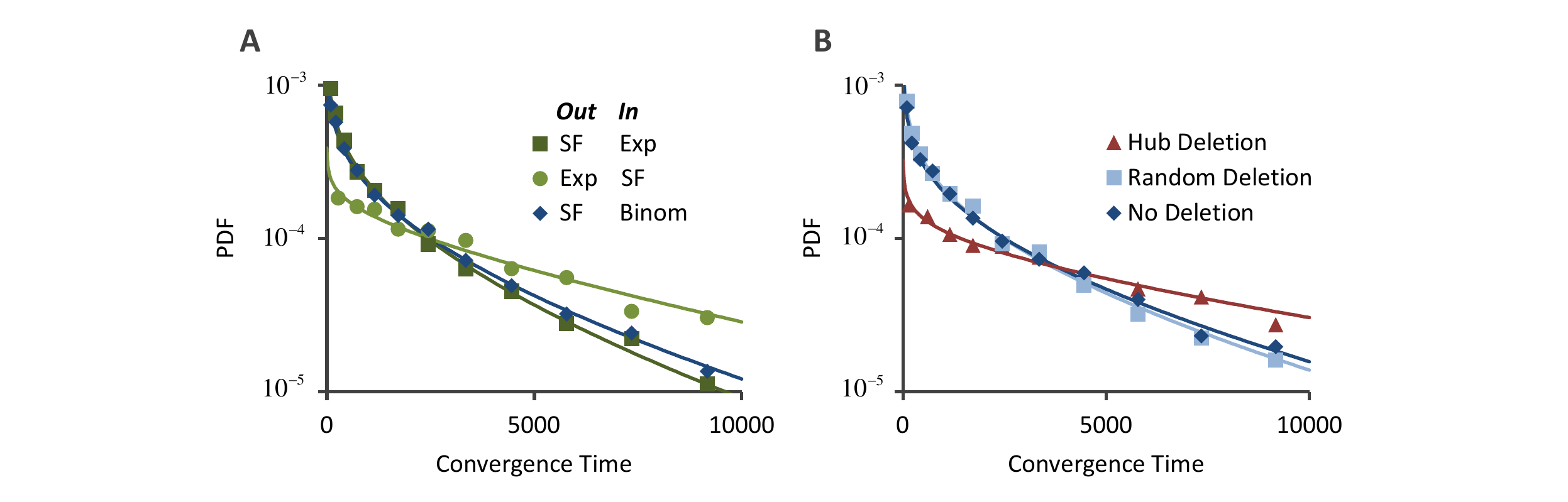}
		\caption{{\bf Distribution of convergence times} for networks which converged in less than $10^4$ timesteps. Solid lines depict stretched exponential fits. {\bf (A)} Probability density distribution (PDF) of convergence time for three topological ensembles. {\bf (B)} PDFs after deleting the 8 largest hubs (red) or the same number of randomly-chosen nodes (light blue) from the SF-Binom ensemble. All ensembles have $N=1000$,  $y*=0$,  $g_0 = 10$,  and  $D=10^{-3}$. Degree distribution parameters: SF, $a=1$, $\gamma=2.4$; Binom, $(p\simeq \dfrac{3.5}{N}$, $N)$; Exp, $\beta \simeq 3.5$.}
		
		\label{fig:Distributions}
	\end{center}
\end{figure*}

\bigskip
\noindent \textbf{\textit{Adaptation Success Correlates with Abundance of Attractors}}

\noindent In the typical example shown in Fig. 1, exploratory dynamics culminates in reduction of drive for exploration and convergence to a stable attractor of Eq. (1). The significant differences between adaptive performance of network ensembles (Fig. 2A,B) may reflect the abundance of networks supporting relaxation to attractors in the different ensembles. 
Previous work has shown that for networks with uniform degree distributions and sufficiently strong interactions, the number of attractors of Eq. (1) decreases with network size and vanishes in the limit of infinite size (leading to chaotic motion only \cite{Sompolinsky88}). A related result was recently found for Boolean networks \cite{Pinho2012most}. It is not known, however, how the number of attractors scales with system size for networks of arbitrary topological structured.

To address this question, we simulated  many independent networks in each ensemble and estimated the fraction which relaxed to fixed points without exploration or feedback (Eq. (1) alone). For any given network the probability of relaxation to a fixed point was found largely insensitive to the initial conditions in x-space (not shown). With that in mind we computed, for each topological ensemble, the fraction of networks supporting relaxation within a given time window, starting with random initial conditions. This measure is analogous to the CF used in Fig. 2, but without a constraint, feedback or random walk in connection strengths. To highlight the dependence on network size we extended the simulations up to $N=10000$. Fig. 7A reveals topology-dependent differences that are qualitatively in line with the ability for exploratory adaption shown above (Fig. 2B). This suggests that a substantial contribution to successful adaptation is indeed provided by a high abundance of networks exhibiting fixed points in their dynamics. 

\begin{figure*}
	\begin{center}
			\includegraphics[width=16cm]{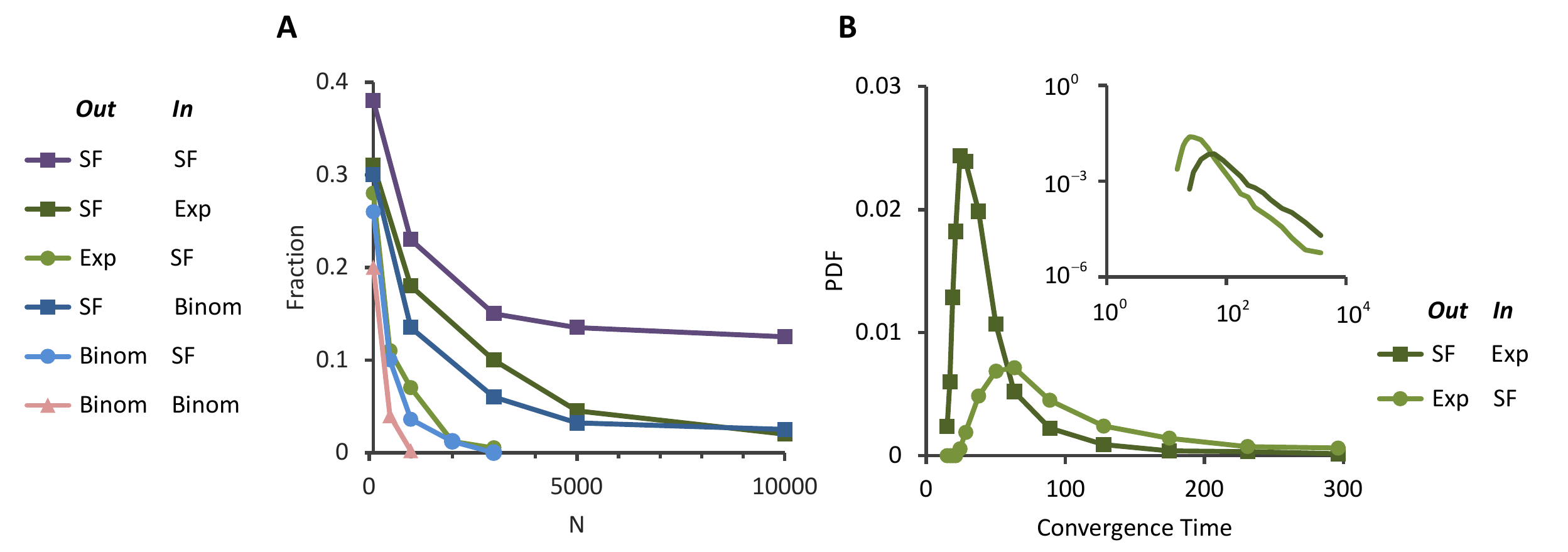}
		
		\caption{ {\bf Fixed points in the absence of exploration for different topological ensembles.} {\bf (A)} Fraction of networks within an ensemble which relaxed to a fixed point under the nonlinear dynamics of Eq. (1), with fixed connections, no constraint and no feedback. Topological ensembles which exhibited higher success in exploratory adaptation in Fig. 2B, relaxed to fixed points in a larger fraction of simulations. {\bf (B)} Distribution of relaxation times to fixed points for two of the ensembles. Note the shorter typical timescale for the SF-Exp ensemble (the more successfully adapting ensemble). $N=1000$, $g = 10$. Dergree distribution parameters: SF, ($a=1$, $\gamma=2.4$); Binom, $(p\simeq \dfrac{3.5}{N}$, $N)$; Exp, $\beta \simeq 3.5$.  }
		
		\label{fig:Distributions}
	\end{center}
\end{figure*}

For each network ensemble that supports fixed points, we further analyzed the distribution of  relaxation times into these fixed points. Fig. 7B demonstrates the effect of topology by comparing the SF-Exp ensemble to the transposed Exp-SF. It shows that networks with SF-out degree distribution typically support faster relaxation to their respective fixed points. This may allow the adaptation to converge before exploration has had a chance to significantly modify network connections. 
Further work is required to test this hypothesis and to broaden the theoretical understanding of these dynamics in random ensembles with heterogeneous topology.


\bigskip
{\fontsize{14}{14} {\noindent \bf Discussion}}

\noindent Overall, we have introduced a model of exploratory adaptation driven by mismatch between an internal global variable and an external constraint. 
Adaptation is achieved by a purely exploratory process which relies on the plasticity of regulatory interactions \cite{Niklas2015,Bondos2016flexibility}. Our model was formulated in terms of gene regulation but other cellular interactions, such as protein-protein interaction networks, may also constribute to similar adpatation. We have found that convergence of exploratory adaptation depends crucially on structural properties of the network. It requires the existence of outgoing hubs and is enhanced by auto-regulation of these hubs. These results offer an important, but hitherto unrealized, rationale for the overwhelming abundance of autoregulation motifs on master regulatory transcription factors \cite{Pinho2014stability}. These master regulators act as network hubs by virtue of the large numbers of their downstream gene targets. Our findings show that autoregulation of such hubs dramatically improves their ability to drive the network into a stable state which satisfies a phenotypic demand.

The contribution of outgoing hubs to the success of adaptation may reflect their ability to coordinate changes in a large set of affected nodes. In a network with a narrow distribution of out-degrees (without hubs), each node has the same relatively small influence as any other node. In the absence of a hierarchy in the extent of influence, irregular dynamic variation
is unlikely to sum into a coherent change in the phenotype. On the other hand, the existence of a few hubs with a much broader influence can promote correlations between many downstream nodes, leading to an increase in the ability to drive a coherent change in a given direction. These effects may be related to other aspects of stability in network dynamics that vary with topology \cite{Aldana2003,Hazan2012,deEspanés2016}.

Beyond the structural aspects promoting exploratory adaptation, the process of convergence itself appears to be complex and is characterized by an extremely broad distribution of times. Successful convergence likely depends on a delicate interplay between the space of possible network configurations, their connectivity properties
and the typical timescales of their intrinsic dynamics.

While our model draws from neural network models \cite{Maass2002,SussilloAbbott09,Barak2013}, it is substantially different in relying on purely stochastic exploration. In the language of learning theory, the "task" is modest: convergence to a stable attractor which satisfies a low-dimensional approximate constraint. Without exploration, this task could be fulfilled by chance with a very small probability. This probability increases dramatically by exploratory dynamics within a class of networks of a given structure. The ability to achieve high success rates without a need for complex computation or fine-tuning  makes this type of adaptation particularly plausible for biological implementation. The relevance of similar processes in neural networks remains to be investigated.

Random-network models were previously used to address evolutionary dynamics of gene regulation over many generations. These studies considered a population of networks undergoing random mutations and selection according to an assigned fitness \cite{Wagner2011,BergmanSiegal2003}. In contrast, the model presented in the current study considers random variations over time within a single network, as an abstraction of a particular aspect of single cell adaptation within its lifetime. While these two approaches differ in timescales, level of organization and biological phenomena, it seems that they cannot be completely decoupled and that biological networks have basic properties that reflect on both contexts \cite{Barzel2013}. For example, marked differences in evolutionary dynamics were found between homogeneous and SF networks \cite{OikonomouCluzel}. 
In fact, the reproducible and exploratory responses in single cells, and the evolutionary processes at the population-level, correspond to complementary aspects of gene-environment interactions at different scales \cite{LopezMaury2008,Pilpel2015}. A major future goal would be to integrate these aspects into a general picture of adaptive responses to diverse types of challenges over  a broad range of timescales. 


\bigskip
\bigskip
\fontsize{14}{14} {\noindent \bf Methods}
\large

\small {

{\noindent \bf Constructing Network Backbone {\bf $T$} For Topological Ensembles} \\
Interactions between the intracellular dynamical variables are governed by the network matrix $W$, defined as the element wise (Hadamrd) product of the binary backbone, the adjacency matrix $T$, and a Gaussian random matrix $J$ of connection strengths (Eq. 2). We construct an ensemble of given topology by sampling the connectivities of the backbone from given in-degree and out-degree distributions,  $P_{in}(K^{in})$ and $P_{out}(K^{out})$, and by sampling the random strengths of $J$ independently from a Gaussian distribution.
In practice, $T$ is constructed first by randomly sampling a list of $N$ out-going degrees $\left\lbrace{d^{out}_i}\right\rbrace_{i=1}^{N}$ from the distribution $P_{out}(K^{out})$ with $d^{out}_i \leqslant N-1$; and then sampling a list of $N$ in-coming degrees $\left\lbrace{d^{in}_i}\right\rbrace $ from the distribution $P_{in}(K^{in})$ (again $d^{in}_i \leqslant N-1$), conditioned on the graphicality of the in- and out- degree sequences \cite{Chartrand1986}. The network is then constructed from these sequences using the algorithm described in \cite{kim2012}.
\par Scale-Free (SF) sequences are obtained by a discretization to the nearest integer of the continuous Pareto distribution $P(K) = \dfrac{(\gamma-1)a^{(\gamma-1)}}{K^\gamma}$. Sampling SF degree sequences using the discrete Zeta distribution gives qualitatively similar results (results not shown). Binomial sequences are drawn from a Binomial distribution $P(K) =\mathcal{B}\bigl( N, p\bigr)$, with $p=\dfrac{\bra K \ket}{N}$. Exponential sequences are obtained by a discretization to the nearest integer of the continuous exponential distribution $P(k) = \dfrac{1}{\beta}e^{-\dfrac{K}{\beta}} $ with $\beta = \bra K \ket$. A Binomial degree sequence is implemented using MATLAB built-in Binomial random number generator. Exponential and Scale-free sequences are implemented by a discretization of the continuous MATLAB built-in Exponential and Generalized Pareto random number generators with Generalized Pareto parameters $k=1\big/({\gamma-1})$, $\sigma=a\big/({\gamma-1})$ and $\theta=a$. 

\bigskip 
{\noindent \bf Comparison Between Different Ensembles} \\
To compare adaptation performance between different ensembles, interaction matrices need to be properly normalized. In the study of uniform random matrices, the elements are usually normalized such that their variance is $\dfrac{{g_0}^2}{N}$, providing a well-defined thermodynamical limit $N \rightarrow \infty$ in which the matrix eigenvalues of are uniformly distributed within a disc of size $g_0$ in the complex plane \cite{Sommers88, Wood2012}. 

In our model the interaction matrix is a product of a topological backbone, the binary adjacency matrix $T$, and a interaction strength matrix $J$. The initial interaction matrix $J_0 \triangleq J(t\!=\!0)$ is defined as a random Gaussian matrix with mean $0$ and variance $\dfrac{{g_0}^2}{\bra K\ket}$, ${\bra K\ket}$ being the average connectivity. Neglecting correlations in the adjacency matrix $T$, the variance of its elements is $Var(T_{uj})=\dfrac{\bra K \ket}{N} \left(1-\dfrac{\bra K \ket}{N}\right)\approx \dfrac{\bra K \ket}{N}$, which implies  $Var(W_{i,j}) \approx \dfrac{{g_0}^2}{N}$. 
In principle both finite-size effects and correlations in $W_{i,j}$ result in deviations from a uniform distribution of eigenvalues in the circle. However empirically we find that for matrices of relevant size, the spectral radius of $W$ is still $\sim g_0$, establishing a basis for comparison between the different ensembles based on spectral radius. We note however that the eigenvalue distribution is far from being uniform (see Supplementary note 1, Empirical spectrum of interaction matrices $W$). 

Another model component that needs to be normalized for proper comparison is the macroscopic phenotype $y(\bm {\mathbf{x}})=\bm {\mathbf{b}}\cdot \bm {\mathbf{x}}$. The arbitrary weight vector $\bm {\mathbf{b}}$ is characterized by a degree of sparseness $c$, i.e. the fraction of nonzero components, $\dfrac{1}{N}\!\!<c\!<1$; and by the typical magnitude of those nonzero components. In order to compare between networks of different sizes and weight vectors of different sparseness, the variance  of the non-zero components is scaled by their number, $cN$ and by the matrix gain ${g_0}^2$. The non-zero components of $\mathbf{b}$ are thus distributed $b_{i} \sim \mathcal{N}(0,  \dfrac{1}{{g_0}^2\cdot cN}\cdot\alpha)$, with $\alpha$ a single parameter that determines the typical scale of the phenotype fluctuations in different network sizes and gains (See Supplementary Note 1, Distributions of phenotype $y$).

\bigskip

{\noindent \bf Computing Convergence Fractions}\\
Convergence fractions were computed over 2000 time steps in samples of 500 networks drawn from specified in- and out-degree distributions, averaging over $T$, $J_0$ and $\mathbf{x}_0$. For  fully or sparsely connected homogeneous random networks of size $N=1500$, the CF is close to zero (not shown). Alternative ensemble definitions (e.g. keeping $T$ fixed) do not change the main results (see Supplementary Note 1, Convergence of different network ensembles).

\bigskip

{\noindent \bf Saturating function $\phi( \mathbf{x})$} \\
The saturating function $\phi(\bm {\mathbf{x}})$ is defined as an element-wise function $\phi(x_j)= \tanh (x_j)$ operating separately on each of the components of $\bm {\mathbf{x}}$. Model results are insensitive to the exact shape of this function 
(Supplementary Note 2, Robustness of model to saturating function $\phi$) and to placing the saturation inside or outside of the interactions (Supplementary Note 2, Robustness of model to position of saturating function $\phi$).

\bigskip

{\noindent \bf Mismatch function ${\cal M}(y)$}\\ 
The mismatch function ${\cal M}(y)$ is defined here as ${\cal M}(y) = \dfrac{{\cal M}_0}{2}\left[1+\tanh\left(\dfrac{|y-y^*|-\varepsilon}{\mu}\right)\right]$, a symmetric sigmoid around $y^*$, where $\varepsilon=3$ controls the size of the low-mismatch "comfort-zone" around $y^*$, $\mu=0.01$  the steepness of the sigmoid, and ${\cal M}_0=2$ its maximal value. Main model results are insensitive to the exact shape of this function as long as it has a flat region with zero or very low mismatch around $y^*$.  (see Supplementary Note 2, Robustness of model to mismatch function ${\cal M}$).

\bigskip
\bigskip

\bibliographystyle{unsrt}

	{\noindent \bf Supplementary Information} to this manuscript is provided as a separate document.
	\vspace{14 pt}
	
	{\noindent \bf Acknowledgments} We thank O. Barak, E.Braun, R. Meir, and M. Stern for valuable discussions and  S. Marom, A. Rivkind, L. Geyrhofer and H. Keren for critical reading of the manuscript. 
	\vspace{14 pt}
	
	{\noindent \bf Authors Contributions}  
	Y.S. and N.B conceived the general approach for modelling adaption by exploratory dynamics. H.S. and N.B. constructed the model. H.S. performed all the simulations and computations. All authors evaluated model findings and designed simulations to identify requirements and properties of exploratory adaptation. All authors wrote the manuscript.
	\vspace{14 pt}
	
	{\noindent \bf Author Information} The authors declare no competing financial interests.
	Correspondence and material requests should be addressed to N.B. (nbrenner@technion.ac.il).
	\vspace{14 pt}
	
	 {\noindent \bf Data Availability}
	 The data that support the findings of this study are available from the corresponding
	author upon reasonable request.
}

\end{multicols}

\newpage  
 
\setcounter{page}{1}
\setcounter{equation}{0}
\renewcommand{\thefigure}{A\arabic{figure}}
\setcounter{figure}{0}   
\section*{\centering \huge {Supplementary Note 1}  }
\vspace{10ex}

\subsection*{Empirical Spectrum of Interaction Matrices $W$}

The initial interaction matrix $J_0 \triangleq J(t=o)$ is defined as a random Gaussian matrix with mean $0$ and variance $\dfrac{{g_0}^2}{\bra K\ket}$, where  $\bra K \ket$ is the mean in and out degree, and
$g_0$ (the network gain) determines the spectral radius of the combined interaction matrix $W$ at $t=0$ (see Methods section in main text). 
Empirically we find that for matrices of relevant size the spectral radius of $W$ is not greatly affected by topology and it remains $\sim g_0$ following the above normalization $Var({J_0}_{ij})=\dfrac{{g_0}^2}{\bra K\ket}$ , however the distribution is highly non-uniform (Sup. Fig. 1) 

\begin{figure}[H]
	\refstepcounter{figure}
	\label{fig:eigen}
	\begin{center}
		\includegraphics[width=16cm]{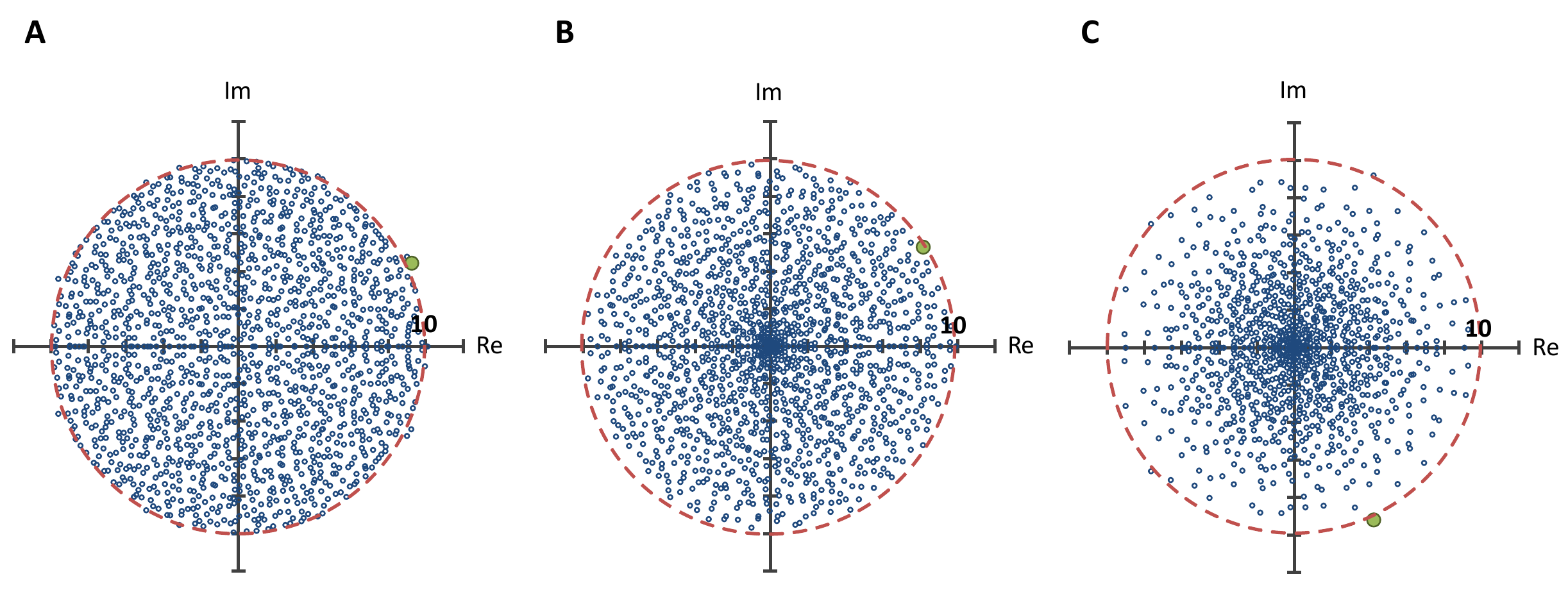}
	\end{center}
	\renewcommand{\baselinestretch}{1}
	\small{
		\textbf{Supplementary Figure \arabic{figure}. Eigenvalues of finite size matrices with N=1500}. The eigenvalues of full Gaussian (A), sparse Gaussian (B) and scale-free/binomial (C) matrices  are plotted. The eigenvalues of all three matrices are almost entirely contained within a disc of radius $g_0$ (broken red) and all three have a largest norm of eigenvalue $\sim g_0$ (green dot). However, the distribution of eigenvalues in the disc differs considerably between the three matrices.  The number of non-zero elements in the sparse Gaussian matrix (B) is distributed with Binomial distribution  in both columns and rows. The scale-free/binomial matrix (C) has a Binomial distribution for the number of non-zero elements in the rows and a scale-free distribution in the columns. All matrices have the form $W=T\circ J$, with  ${J}_{ij} \sim \mathcal{N}\bigl( 0, \dfrac{{g_0}^2}{\bra K\ket}\bigr)$ and $g_0=10$.    Binomial distributions in (B) and (C) have parameters $p\simeq \dfrac{5}{N}$ and  scale-free distribution in (C) has parameters $a=1$, $\gamma = 2.2$.}
	\label{fig:Spectra}	
\end{figure}

\subsection*{Distributions of Phenotype $y$} 
The variable representing the macroscopic phenotype is defined as $y(\mathbf{x})=\mathbf{b}\cdot \mathbf{x}$. The arbitrary weight vector $\mathbf{b}$ is characterized by a degree of sparseness $c$, i.e. the fraction of nonzero components, $\dfrac{1}{N}\!\!<c\!<1$. The non-zero components of $\mathbf{b}$ are thus distributed $b_{i} \sim \mathcal{N}(0,  \dfrac{1}{{g_0}^2\cdot cN}\cdot\alpha)$ (see Methods section in main text). Sup. Fig. 
2(A-C) depicts distributions of the values of $y$ for $\alpha = 100$ with various types of interaction matrices $W$. As can be seen, these distributions  are similarly shaped for a broad range of network sizes (Sup. Fig. 
2A) and gains, $g_0$ (Sup. Fig. 
2B), and do not change for various network topologies (Sup. Fig. 
2C). These results verify that $y$ and $J_0$ are appropriately normalized. 
\begin{figure}[H]
	\refstepcounter{figure}
	\label{fig:eigen}
	\begin{center}
		\includegraphics[width=16cm]{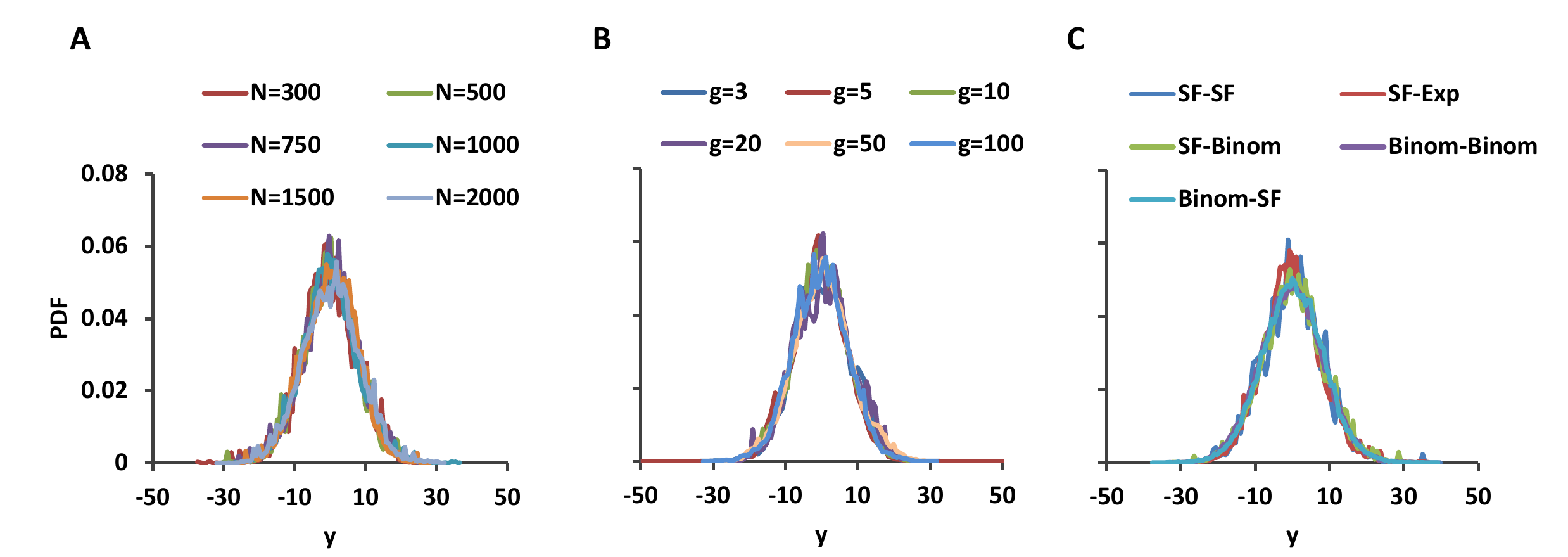}
	\end{center}
	\renewcommand{\baselinestretch}{1}
	\small{
		\textbf{Supplementary Figure \arabic{figure}. Distributions of phenotype $y$ over trajectories for various network ensembles} The distribution of the values of the macroscopic phenotype $y=\mathbf{b}\cdot\mathbf{x}$  are plotted for various ensembles of networks with fixed interaction strengths. These include networks of various sizes (A), network gains (B) and topologies (C). For all ensembles the $y$ values are similarly distributed. This indicates that $y$ and $J_0$ are appropriately normalized. For all networks $\alpha = 100$. Networks in (A) and (B) have Scale-free out-degree distribution and Binomial in-degree distribution; Networks in (A) and (C) have $g = 10$ and networks in (B) and (C) have $N=1000$. In all panels scale-free in/out distributions have parameters $a=1$ and $\gamma=2.4$, exponential distributions have parameter $\beta =3.5$ and binomial distributions have parameters $p= \dfrac{3.5}{N}$ and $N$. }
	\label{fig:YNorm}	 
\end{figure}

\subsection*{Convergence of Different Network Ensembles}
The topological ensembles in our model includes both quenched and annealed disorder. The random topology of the network, namely the specific adjacency matrix $T$ , is quenched and remains the same throughout the course of any single simulation run. The strengths of the network interactions, $J(t$), on the other hand, are dynamic and change via a random walk, thus presenting an annealed disorder. Convergence fractions are computed by averaging over such simulations; one needs to determine what is the relevant ensemble to average over. 
\par Given a choice of the model parameters, one possible ensemble $\{(T^j,J_0^j, \mathbf{x}_0^j)\}_{j=1}^{m}$, consists of a set of $m$ networks, each with a different topology $T^j$, different initial interaction strengths $J_0^j \triangleq J^j(t=0)$ and different initial conditions $\mathbf{x}_0^j \triangleq \mathbf{x}^j(t=0)$. Another potential ensemble, $\{(T^0,J_0^j, \mathbf{x}_0^j)\}_{j=1}^{m}$, contains of a set of networks which all share the same adjacency matrix $T^0$, but differ in the initial network strengths $J_0^j$, and initial conditions  $\mathbf{x}_0^0$; A third possibility is constructing an ensemble by varying only the initial conditions $\mathbf{x}_0^j$ and using the same initial network $W^0 =T^0\circ J_0^0$, $\{(T^0,J_0^0, \mathbf{x}_0^j)\}_{j=1}^{m}$, and finally, one can simulate the dynamics consecutively keeping both the initial network $W^0$ and initial dynamical conditions $\mathbf{x}_0^0$ constant, $\{(T^0,J_0^0, \mathbf{x}_0^0)\}_{j=1}^{m}$, with different realizations of the exploration process. Whether or not these various ensembles show qualitatively similar statistical properties or not is \textit{a-priori} known and depends on the self-averaging properties of the system.  
\par We tested these properties by computing the distribution of convergence times for the various ensembles. Sup Fig. 
3 shows that these distributions are similarly shaped for all ensembles.

\begin{figure}[H]
	\refstepcounter{figure}
	\label{fig:eigen}
	\begin{center}
		\includegraphics[width=16cm]{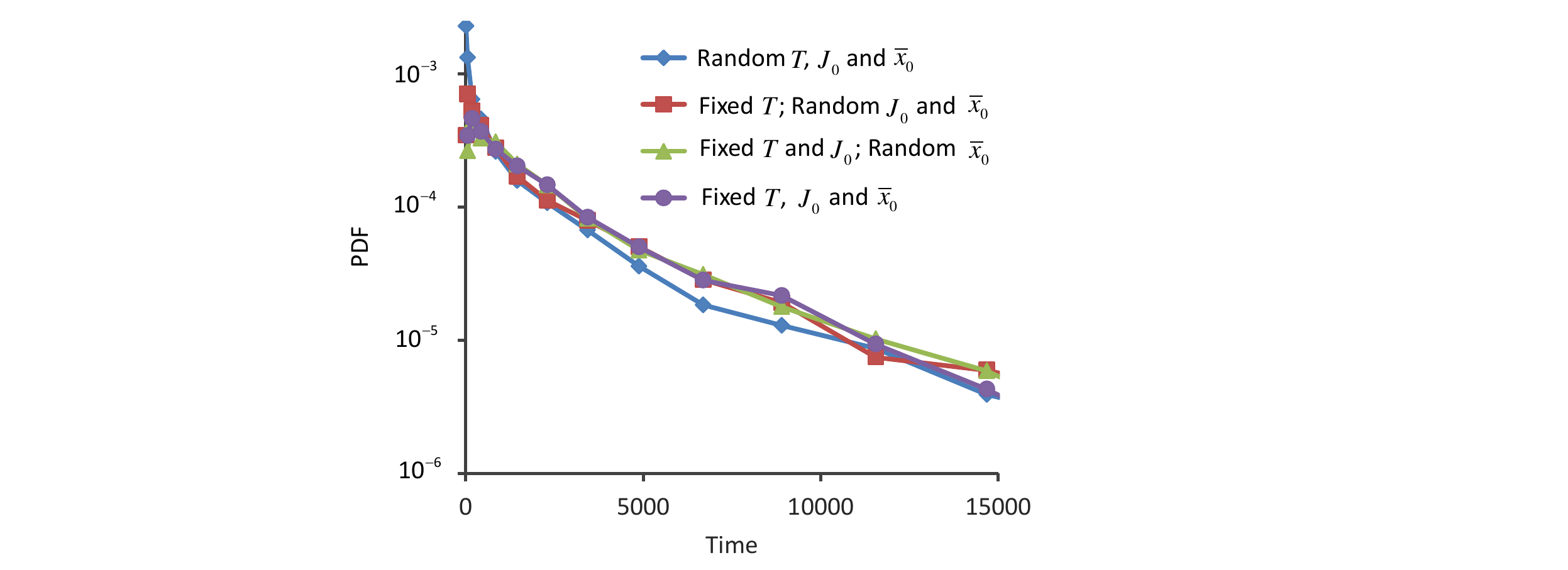}
	\end{center}
	\renewcommand{\baselinestretch}{1}
	\small{
		\textbf{Supplementary Figure \arabic{figure} Convergence Time distributions for Different Network Ensembles.} (i) An ensemble in which each network has random $T^j$, $J_0^j$ and  $\mathbf{x}_0^j$ (blue); (ii) An ensemble in which all networks share the same topology $T^0$, but differ in $J_0^j$, and $\mathbf{x}_0^j$ (red); (iii) An ensemble in which initial network $W =T^0\circ J_0^0$ is the the same for all networks but initial conditions $\mathbf{x}_0^j$ are unique (green) (iv) An ensemble in which both the initial network $W^0$ and the initial dynamical conditions $\mathbf{x}_0^0$ are the same for all networks. All Ensembles have SF out-degree distribution and Binomial in-degree distribution. The backbone $T$ is the same matrix for ensembles (ii), (iii) and (iv) and the initial interactions strengths $J_0$ is the same in ensembles (iii) and (iv). For all ensembles $N=1500$, $g_0 = 10$, $\alpha = 100$, $\mathcal M_0 = 2$, $c=0.2$, $\varepsilon = 3$, $D=10^{-3}$ and $y*=0$. SF out-degree distribution has parameters $a=1$, $\gamma=2.2$, and Binomial in-degree distributions has parameters $p= \dfrac{5}{N}$ and $N$.}
	\label{fig:EnsembleAveraging}	
\end{figure}  

\newpage  
\section*{\centering \huge {Supplementary Note 2}  }
\vspace{10ex}
\setcounter{equation}{0}
\subsection*{Robustness of Model to Saturating Function $\phi$}

The dynamics of the microscopic variables  $\mathbf{x}$ prior to any exploration in $W$ is given by 
\be
\dot{ \mathbf{x}}&=&W\phi(\mathbf{x})-\mathbf{x}.
\ee
The results shown in the main text were obtained using the element-wise saturating function $\phi(x_i) = \tanh(x_i)$. However, we find that these main results hold also for other types of saturating functions, specifically piece-wise linear and Sign function. In all cases convergence fractions depend on the topology of the networks, with higher fractions for those with scale-free out-degree distribution (Sup. Fig. 4A,B). 
The slope of the saturating function at zero has little impact on convergence fractions (Sup. Fig. 4C,D)
\begin{figure}[H]
	\refstepcounter{figure}
	\begin{center}
		\includegraphics[width=13cm]{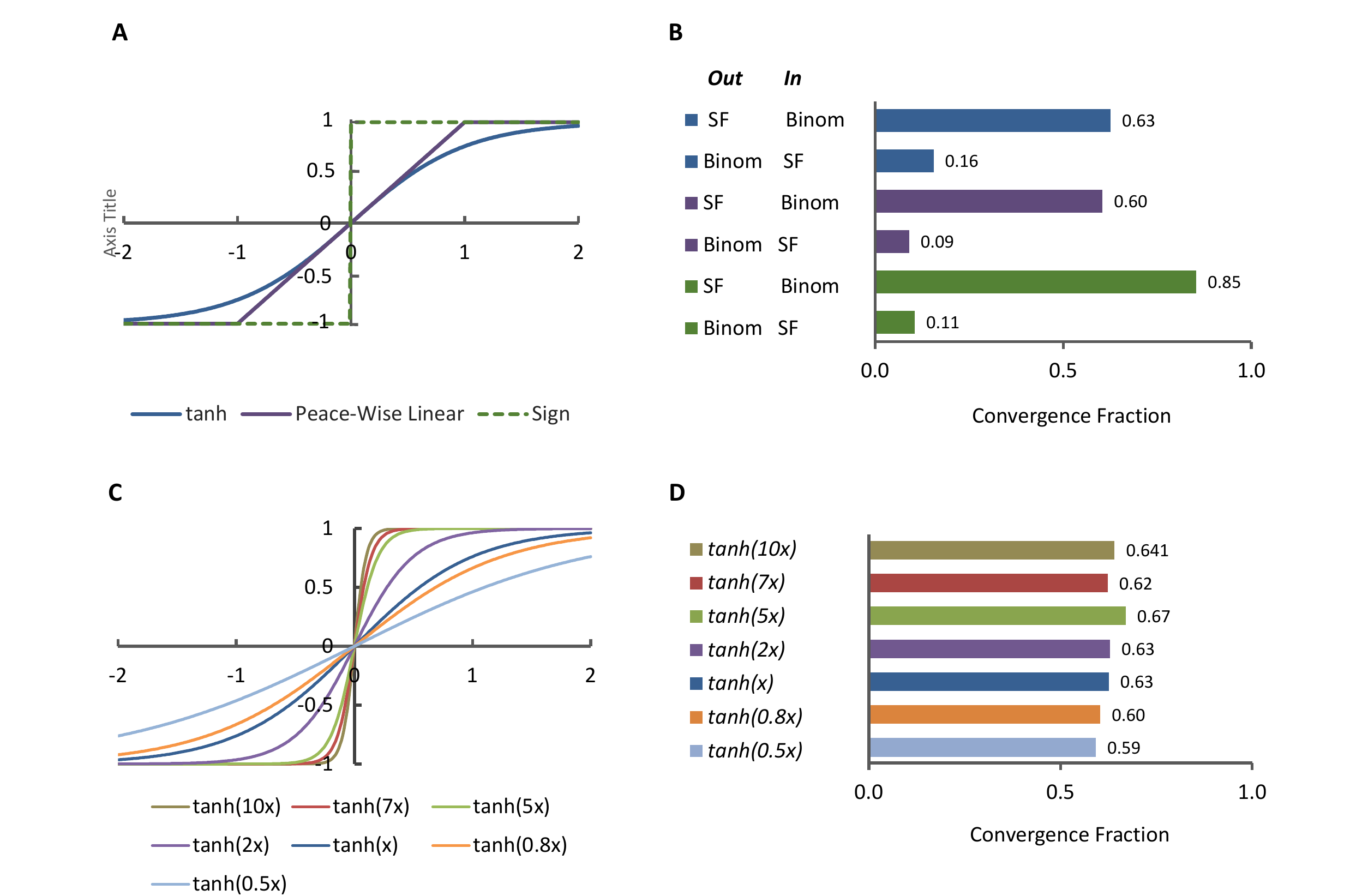}
	\end{center}
	\renewcommand{\baselinestretch}{1}
	\small{
		\textbf{Supplementary Figure \arabic{figure}.  Convergence fractions for dfferent saturating functions.} (A) Functional from of three saturating functions examined: $\tanh(x_i)$ (blue), piece-wise linear (purple) and Sign function (broken green)  (B) Convergence fractions for ensembles with the three functional forms and two types of topology SF-Binom and Binom-SF within a time window of 2000 units .(C) Functional from of saturating functions with various slops at zero. (D) Convergence fractions for ensembles with functional forms of $\phi$ shown in (C) with SF-Binom topology, within a time window of 2000 units. The ensemble samples in (B) and (D) consist of 500 networks each. For all networks $g_0 = 10$, $\alpha = 100$, $c=0.2$ $\mathcal M_0 = 2$, $D=10^{-3}$ and $\varepsilon = 3$,  $y*=0$. Scale-free in/out distributions have parameters $a=1$ and $\gamma=2.4$,  $\beta \sim 3.5$ and Binomial distributions have parameters $p\simeq \dfrac{3.5}{N}$ and $N=1000$. }
	\label{fig:SatFn}	
\end{figure}

\subsection*{Robustness of Model to Position of Saturating Function $\phi$}

In the model described in the main text the saturating function $\phi(x_j)$ operates directly on $x_i$ prior to the interactions, while the interactions $w_{ij}$ multiply $\phi(x_j)$ (See Eq. 1 above). However, it is of interest to examine a possible alternative model in which the saturating function $\phi$ operates on $W\mathbf{x}$ and the equation of motion is
\be
\dot{ \mathbf{x}}&=&\phi(W\mathbf{x})-\mathbf{x}.
\ee
or equivalently
\be
\dot{x_i}&=&\phi\left(\sum\limits_{j} Wx_j\right)-x_i.
\ee

Similar equations are often used to describe the dynamics of neural networks, as well as gene interactions.
It is not \textit{a-priori} whether these two formulations will result in similar convergence properties in the context of the  exploratory adaption protocol described here. Remarkably, we find convergence fractions of the two models to be almost identical (Sup. Fig. 5), as long as the macroscopic phenotype $y$ is appropriately normalized (see Sup. Fig. 2).
Recall that for the model described in the main text the elements of the vector $\mathbf{b}$ which defines the phenotype $y$ are given by $b_{i} \sim \mathcal{N}(0,  \dfrac{1}{{g_0}^2\cdot cN}\cdot\alpha)$. The normalizing factor $\dfrac{1}{{g_0}^2\cdot cN}$ ensures $y \sim \mathcal{N}(0,\alpha)$ prior to convergence. The variance of $b_i$ is normalized  by $\dfrac{1}{{g_0}^2}$  due to the empirical distribution of $x_i$ prior to convergence: $x_i  \sim \mathcal{N}(0,\sim {g_0}^2)$. In the alternative model described by Eqs. 2,3 this is not the case. For $g_0 >> 1$,  $x_i$ mostly attains the saturated values of $\phi$ which are $\pm 1$ with equal probability and $Var(x_i)\sim 1$. Thus the normalizing factor $\dfrac{1}{{g_0}^2}$ can be dropped and $b_{i} \sim \mathcal{N}(0,  \dfrac{1}{cN}\cdot\alpha)$ results in $y \sim \mathcal{N}(0,\alpha)$ as in the former case.

\begin{figure}[H]
	\refstepcounter{figure}
	
	\begin{center}
		\includegraphics[width=16cm]{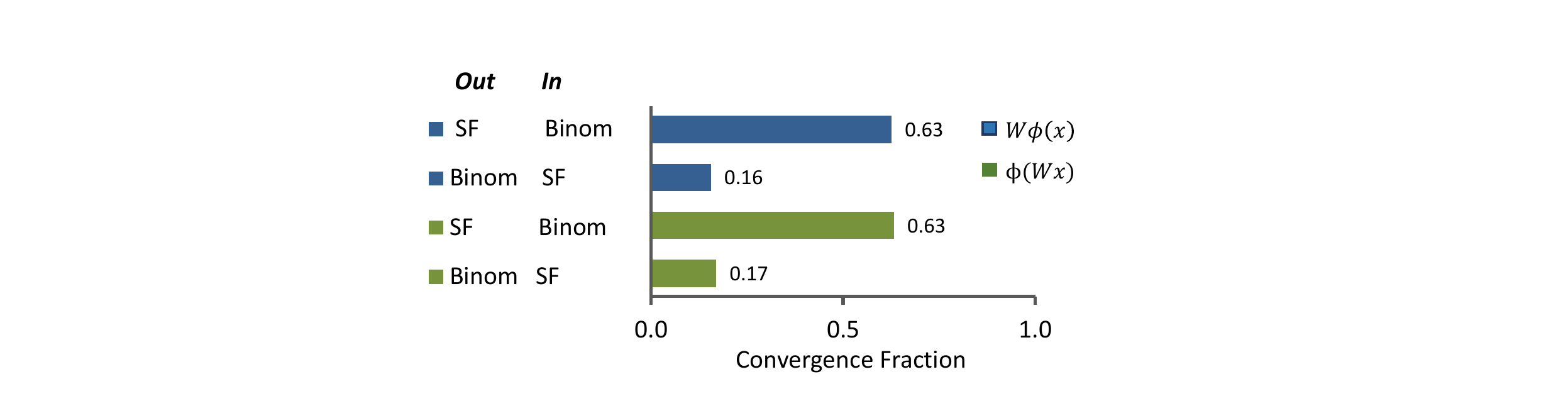}
	\end{center}
	\renewcommand{\baselinestretch}{1}
	\small{
		\textbf{Supplementary Figure \arabic{figure}.  Convergence Fractions with saturating functions inside and outside the summation.}  Convergence fractions within a time window of 2000 units for the model described in the main text (Blue) and a similar model in which the saturating function is placed outside the summation (Green).  Convergence Fractions are shown for ensembles with two types of topology: SF-Binom (out-in) and Binom-SF (out-in) for both models. For all networks $g_0 = 10$, $\alpha = 100$, $c=0.2$ $\mathcal M_0 = 2$, $D=10^{-3}$ and $\varepsilon = 3$,  $y*=0$. Scale-free in/out distributions have parameters $a=1$ and $\gamma=2.4$,  $\beta \sim 3.5$ and Binomial distributions have parameters $p\simeq \dfrac{3.5}{N}$ and $N=1000$. }
	\label{fig:InsideOut}	
\end{figure}

\subsection*{Robusntess of Model to Mismatch function ${\cal M}(y)$} 
For all computations shown in the main text, the mismatch function ${\cal M}(y)$   is defined as a symmetric sigmoid around $y*$

\be
{\cal M}(y) = \dfrac{{\cal M}_0}{2}\Big[1+\tanh \Big(\dfrac{|y-y*|-\varepsilon}{\mu}\Big)\Big],
\ee
\noindent where $2\varepsilon$ is the size of the low mismatch comfort zone around zero, $\mu$ controls the steepness of the sigmoid in its dynamic range, and ${\cal M}_0$ is its maximal value (see Sup. Fig. 6, blue line). An alternative linear mismatch function
\be
{\cal M}(y)= 
\begin{cases} 
	\hfill |y-y*|- \varepsilon   \hfill & |y-y*|>\varepsilon \\
	\hfill 0 \hfill & |y-y*|\leq \varepsilon \\
\end{cases}
\ee
\noindent was examined (see Sup. Fig. 6, red line), resulting in similar convergence properties.
However, using a parabolic function for the mismatch resulted in poor convergence fractions for the same parameters displayed in the main text. Thus, the existence of a broad region of zero mismatch, rather than a well-defined minimum at a point, seems essential for convergence by exploratory adaptation, but the detailed shape of the function does not seem to have a large impact on the results. 

\begin{figure}[H]
	\refstepcounter{figure}
	\label{fig:eigen}
	\begin{center}
		\includegraphics[width=16cm]{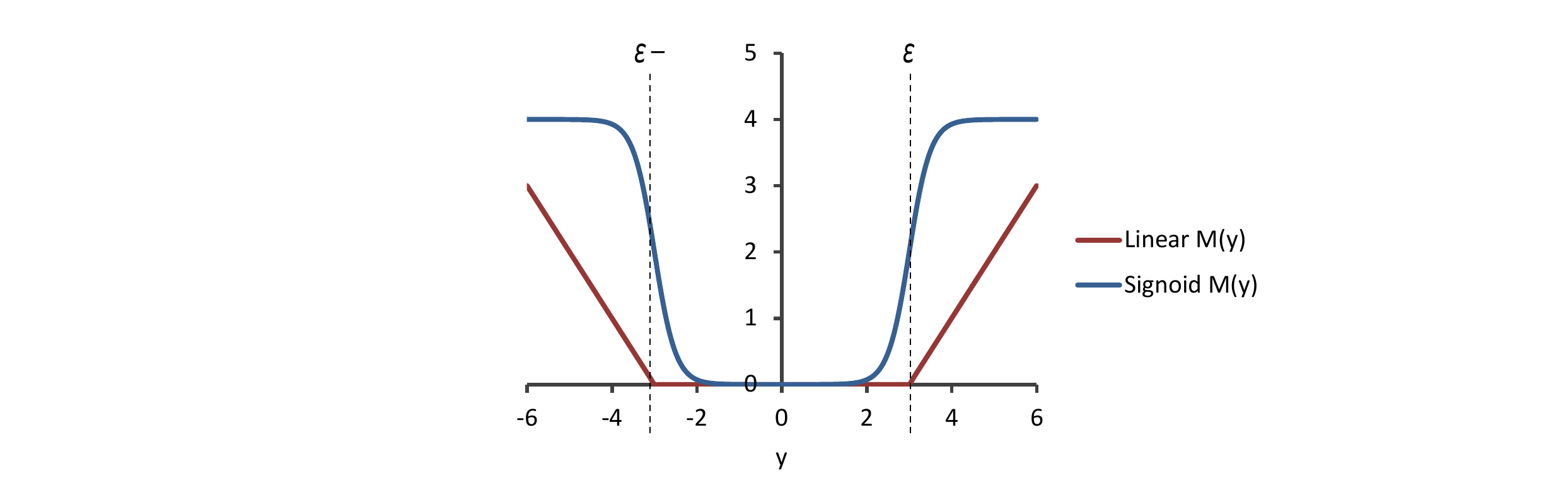}
	\end{center}
	\renewcommand{\baselinestretch}{1}
	\small{
		\textbf{Supplementary Figure \arabic{figure}. Mismatch Functions} Two examples of mismatch functions ${\cal M}(y)$ that result in similar convergence behavior. The results shown in the main text and supplementary were obtained using a sigmoidal mismatch function (blue); similar convergence results can be obtained with a linear mismatch function (red) as well (convergence results not shown). The sigmoidal function in the figure has parameters $\varepsilon = 3$, $mu = 0.5$ and ${\cal M}_0=4$.}
\end{figure}

\newpage
\section*{\centering \huge {Supplementary Note 3}  }
\vspace{10ex}
\setcounter{equation}{0}

\subsection*{Convergence to a Limit Cycle}
An example of convergence to a fixed-point which satisfies the constraint is shown in the main text (Fig. 1 B-D).
However, the non stringent constraint which is reflected in the "comfort zone" of the mismatch function ${\cal{M}}(y)$ allows for a time-varying solutions with small amplitude which are not fixed-points.   
Indeed, many simulations converge to a limit cycle solution (example shown in Sup. Fig. 7 A,B). Such a solution satisfies the constraint only if the amplitude of macroscopic phenotype $y$ is confined in the range $(-\varepsilon, +\varepsilon)$ (Sup. Fig. 7A). The microscopic variables $x_i$ also converge to limit cycles, but these can vary in amplitude and values (Sup. Fig. 7B).  Interestingly for a broad range of network sizes and different network topologies the ratio between convergence of exploratory dynamics to fixed-points and to limit-cycles is largely preserved. For the ensembles shown in Sup Fig. 7C roughly $35\%$ of the solutions are limit cycles and the rest are fixed-points. 

\begin{figure}[H]
	\refstepcounter{figure}
	\label{fig:eigen}
	\begin{center}
		\includegraphics[width=16cm]{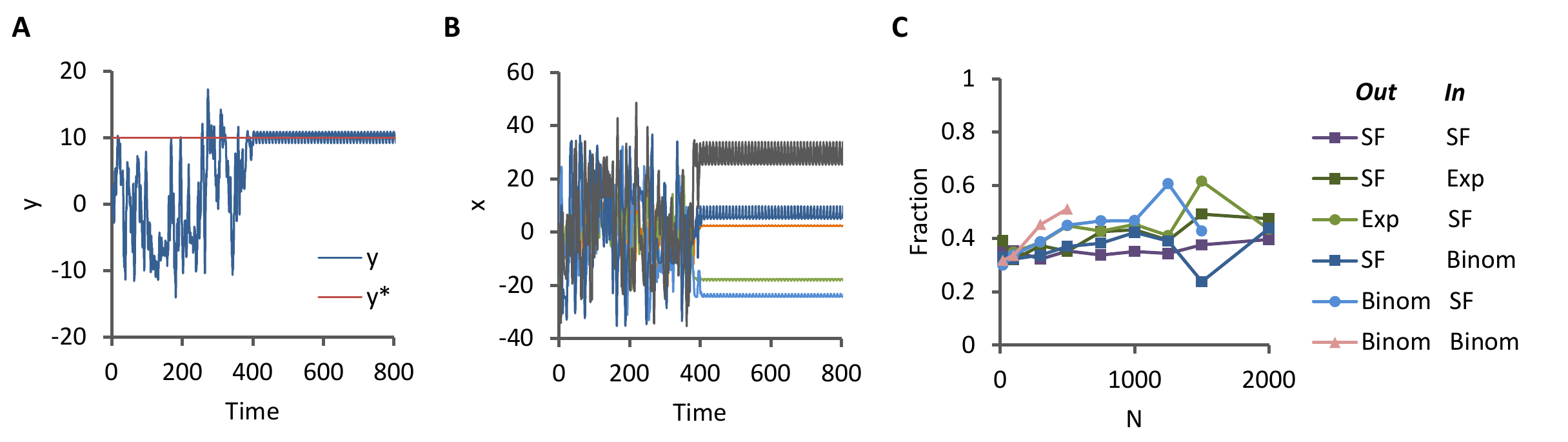}
	\end{center}
	\renewcommand{\baselinestretch}{1}
	\small{
		\textbf{Supplementary Figure \arabic{figure}. Convergence to Limit Cycles.} (A) The macroscopic phenotype $y$ as a function of time in one simulation which converged to a small-amplitude limit tycle around $y^*$. (B) Several microscopic variable $x_i$ as a function of time in the same simulation; $x_i$ also converged to limit cycles but with various amplitudes and centers. (C) Fraction of networks that converged to limit cycles within a time window of 2000 time units, as a function of network size. Results are shown for different ensembles, each composed of a sample of 500 networks. Networks in each of the ensembles has a random $T$, $J_0$ and  $\mathbf{x}_0$. The network in (A) has SF out-degree and Binomial in-degree distributions.  For all networks in (A) (B) and (C) $g_0 = 10$, $\alpha = 100$, $c=0.2$ $\mathcal M_0 = 2$, $D=10^{-3}$ and $\varepsilon = 3$. In (A) and (B) $y*=10$ and in (C) $y*=0$. In all panels scale-free in/out distributions have parameters $a=1$ and $\gamma=2.4$, exponential distributions have parameter $\beta \sim 3.5$ and Binomial distributions have parameters $p\simeq \dfrac{3.5}{N}$ and $N$. }
	
\end{figure}

\subsection*{Dependence of Convergence in Scale-Free Networks on Pareto Distribution Parameters}
As mentioned above, scale-free degree distributions were sampled by discretesizing the continuous Pareto distribution  
\be
P(k) =\dfrac{(\gamma-1)a^{\gamma-1}}{k^\gamma},
\ee
\noindent where the parameter $a$ controls the minimal value of the support and $\gamma$ controls the power law tail of the distribution. In contrast to directly sampling from a discrete distribution such as the Zeta distribution, such a sampling method allows additional control of the lower part of the distribution. After discretization the minimal possible degree, $k_{min}$ is the integer which is nearest to $a$ regardless of non-integer values assigned to $a$. However, the exact value of $a$ affects the weight of the distribution at its minimal value $k_{min}$ and the overall shape of the discrete distribution at its lower part.
For example, for every $a\in [1, 1.5)$ the minimal degree in the network would be 1. However for $a=1.1$ there would be a higher probability for nodes with degree 1 then with $a=1.4$. This allows us to examine with detail the effect of the lower part of the scale-free distribution on the convergence properties of the model.
We find that convergence of exploratory adaptation is indeed sensitive to the the weights at the lower part of the out-going distribution (Sup. Fig. 8A), and occurs with high fractions only for networks with a large enough number of nodes with out-going degree 1 ($a \in (0.4, 1.4)$). In contrast, convergence is weakly dependent on the exact power law of the distribution $\gamma$ (Sup. Fig. 8B).
\par While $a$ and $\gamma$  have a very different effect on the distribution, they both influence the mean degree  $\bra k \ket$ of the network. Sup Fig. 8C shows the same convergence fractions plotted as a function of the mean degree. The results indicate that  $\bra k \ket$ does not  directly influence convergence fractions, and highlights the sensitivity to the lower-part of the distribution which is controlled by $a$. Recent findings have shown that the controllability of random networks is strongly affected by the minimal degree of the nodes with a transition when the minimal degree is increased to $k_{min} \geq 2$ \cite{Menichetti2014}. For our model we conclude that both the existence of hubs and the existence of a large number nodes with out-going degree 1 are indicative of convergence to a stable state.  
\begin{figure}[H]
	\refstepcounter{figure}
	\label{fig:eigen}
	\begin{center}
		\includegraphics[width=16cm]{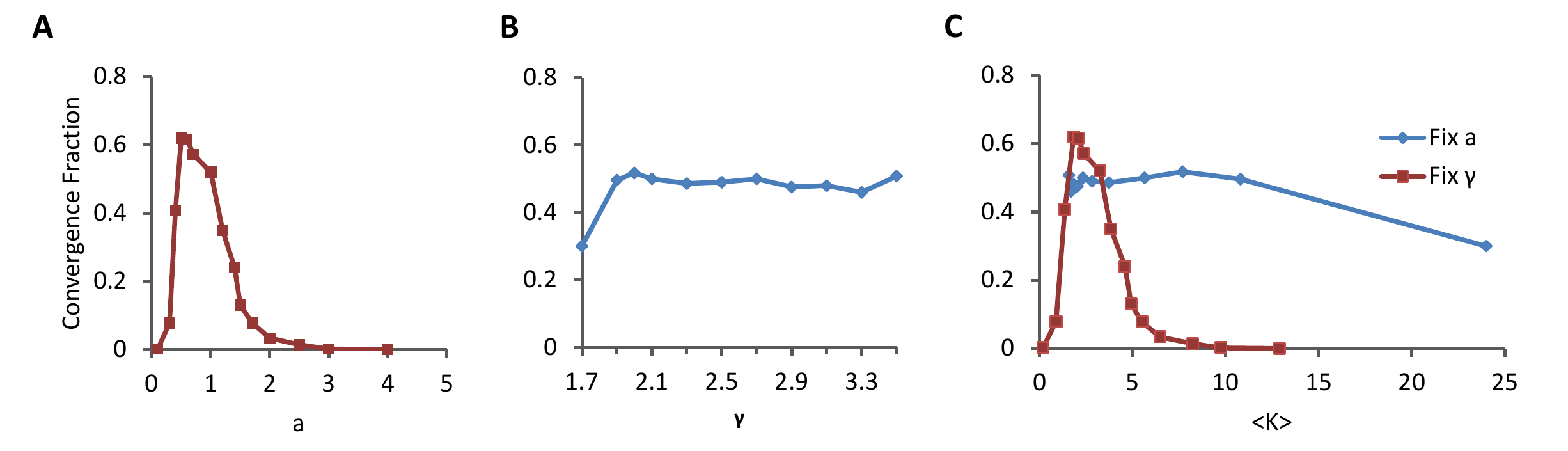}
	\end{center}
	\renewcommand{\baselinestretch}{1}
	\small{
		\textbf{Supplementary Figure \arabic{figure}. Dependence of convergence fractions on parameters of out-degree Pareto Distribution.} Scale-free degree distributions are sampled by discretesizing the continues Pareto distribution  (Eq. 6). (A) Convergence fracion as a function of $a$, a parameter which controls the lower part of the distribution. (B) Convergence fraction as a function of $\gamma$, which controls the power-law tail of the distribution. (C) Convergence fraction as a function of mean degree $\bra K \ket$; changes in this mean degree can be obtained by varying either $a$ (red line) or $\gamma$ (blue line). Each data point in (A), (B) and (C) represents the fraction of network which converged within a time window of 2000 time units from a different ensemble of 500 networks. Networks in each of the ensembles has a random $T$, $J_0$ and  $\mathbf{x}_0$. All Ensembles have SF out-degree distribution and Binomial in-degree distribution. For all results $N=1000$, $g_0 = 10$, $\alpha = 100$, $c=0.2$ $\mathcal M_0 = 2$, $\varepsilon = 3$, $D=10^{-3}$ and $y*=10$. Scale-free out-degree distributions have parameters $a=1$, $\gamma=2.4$, and Binomial in-degree distributions have $p=\dfrac{3.5}{N}$.      }
	
\end{figure}   
\subsection*{Dependence of Convergence on Sparseness of Macroscopic Phenotype}
As mentioned above the macroscopic state, $y(\mathbf{x})=\mathbf{b}\cdot \mathbf{x}$, can have a varying degree of sparseness $c$.
However, we find that convergence properties are not affected by changing the sparseness of macroscopic state (Sup. Fig. 9). This is intuitively understood since the dimensionality of the constraint in the high-dimensional space of microscopic states is the same for all $c$.

\begin{figure}[H]
	\refstepcounter{figure}
	\label{fig:eigen}
	\begin{center}
		\includegraphics[width=13cm]{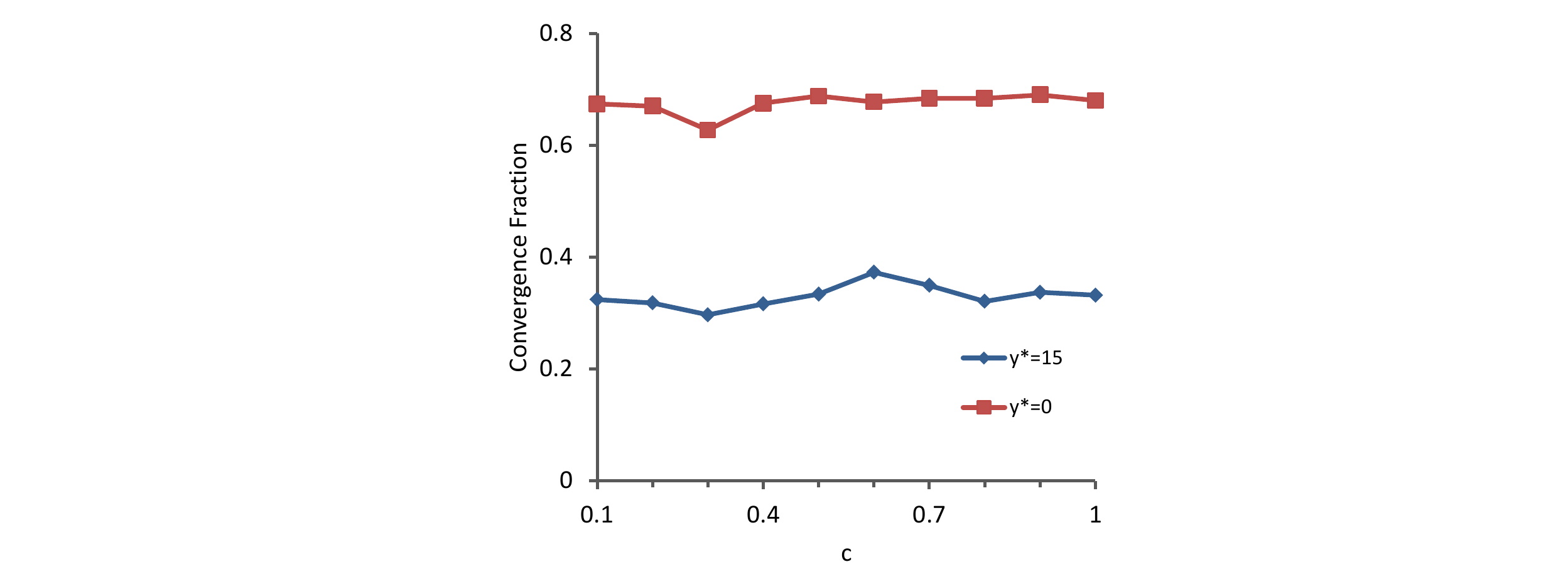}
	\end{center}
	\small{
		\textbf{Supplementary Figure \arabic{figure}. Dependence of convergence fractions on the sparseness of macroscopic state vector.}  Convergence properties are not affected by changing the sparseness of macroscopic state, $c$ (A).  Each data point represents the fraction of network which converged  within a time window of 2000 time units from a different ensemble of 500 networks. Networks in each of the ensembles have random $T$, $J_0$ and  $\mathbf{x}_0$. All Ensembles have SF out-degree distribution and Binomial in-degree distribution. For all results $N=1000$, $g_0 = 10$, $\alpha = 100$, $\mathcal M_0 = 2$ and  $D=10^{-3}$, $\varepsilon = 3$. SF out-degree distribution has parameters $a=1$, $\gamma=2.4$, and Binomial in-degree distributions has parameters $p\simeq \dfrac{3.5}{N}$ and $N$.}
	
\end{figure}

\subsection*{Dependence of Convergence on Network Motifs}
Network motifs are specific local sub-graphs which are thought to be significantly over-represented in gene regulatory networks. We examined the effect of motifs on convergence by creating an ensemble of networks in which motifs are over-represented and comparing the convergence fractions of the motif-enriched networks to an appropriate null model. The single node motif of auto-regulation was discussed in the main text of the article. In this Supplementary section we shall further discuss this motif as well as motifs of higher order. \par

\bigskip
\noindent
\textbf{\textit{Autoregulation}} \par
\noindent the effect of auto regulation of the hubs and positive auto-regulation to random nodes was discussed in the main text. Here we examined separately the effect of adding negative ($W_{ii}<0$) and positive ($W_{ii}>0$) self connections. We assessed the contributions of this motif by creating an ensemble of 500 random networks of size N=1000 and adding auto regulation randomly to 10\% of the nodes (Sup Fig. 10 dark green and dark blue). Such additions change the in and out degrees of some nodes in the networks and consequently affect the overall in and out degree statistics of the network. This is, in general, expected to affect convergence regardless of the auto-regulatory loops. Therefore, for each enriched network we created a null control which shares the same exact in and out degree sequence but does not include an over-representation of the auto-regulatory motif (Sup Fig. 10 light green and light blue). The control was created by randomly re-connecting out half-stubs to in half-stubs until the network is well mixed (see  \cite{shen2002network}) for details of the half-stubs method). Another control is the convergence fractions of the networks prior to any addition (Sup Fig. 10 gray).
We find that both positive and negative auto regulation increases convergence, yet positive auto regulation has a considerably larger effect.

\begin{figure}[H]
	\refstepcounter{figure}
	\label{fig:eigen}
	\begin{center}
		\includegraphics[width=16cm]{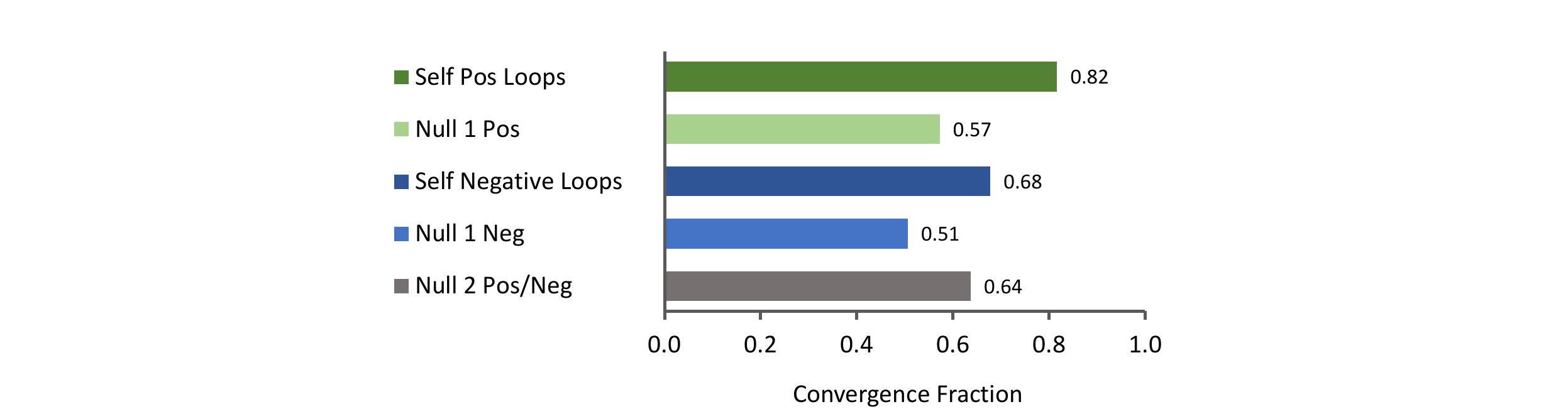}
	\end{center}
	\renewcommand{\baselinestretch}{1}
	\small{
		\textbf{Supplementary Figure \arabic{figure} Effect of adding auto-regulation loops to convergence.}  An ensemble of 500 networks,  in which auto-regulation was added randomly to 10\% of the nodes was tested for convergence. The added connections are either positive (dark green) or negative (dark blue). Results are compared to random networks with the same degree sequence (light green and light blue) and  networks prior to enriching the networks with auto regulation loops (grey). Initial networks prior to the addition of auto regulation loops have SF out-degree distributions with $a=1$, $\gamma=2.4$ and Binomial in-degree distribution with  $p= \dfrac{3.5}{N}$ and $N$. Other parameters are $N=1000$, $g_0 = 10$, $\alpha = 100$, $\mathcal M_0 = 2$, $\varepsilon = 3$, $c=0.2$, $D=10^{-3}$ and $y*=0$. }
	
\end{figure}

\newpage
\noindent
\textbf{\textit{Feed-Forward Loops}} \par
\noindent The 3-node motif which is thought to be most significantly over-represented in regulatory networks is the feed-forward (FF) loop in which $A\Rightarrow B \Rightarrow C$ and $A \Rightarrow C$. We note that the equations governing our model are symmetric around zero and so are the connections strengths $W_{ij}$. Therefore there is no clear interpretation for coherent or incoherent feed-forward loops, and the signs of the connections within each motif were chosen randomly.  We over-represented feed-forward loops in our networks by initially picking a random fraction of existing sequences of the form $A\Rightarrow B \Rightarrow C$ , and adding to these sequences the connection $A \Rightarrow C$ which was not previously part of the network. The strengths of these added connection was drawn from the same distribution as the existing connections in the network. As in the auto-regulation motif we compare these results to a control with the same in and out degree sequence (Sup Fig. 11, blue) and to the original network prior to any addition (Sup Fig. 11, orange).
Networks in the FF ensemble are enriched with 300 feed-forward loops, which increases the over-all number of FF loops by ~50\% on average.

We find that over-representing this motif increases convergence by ~12\% compared to random networks with the same degree sequences (Sup Fig. 11, FF and Null 1). However they do not contribute nor harm the convergence of the network prior to adding the loops (Sup Fig. 11, FF and Null 2).  

\begin{figure}[H]
	\refstepcounter{figure}
	\label{fig:eigen}
	\begin{center}
		\includegraphics[width=16cm]{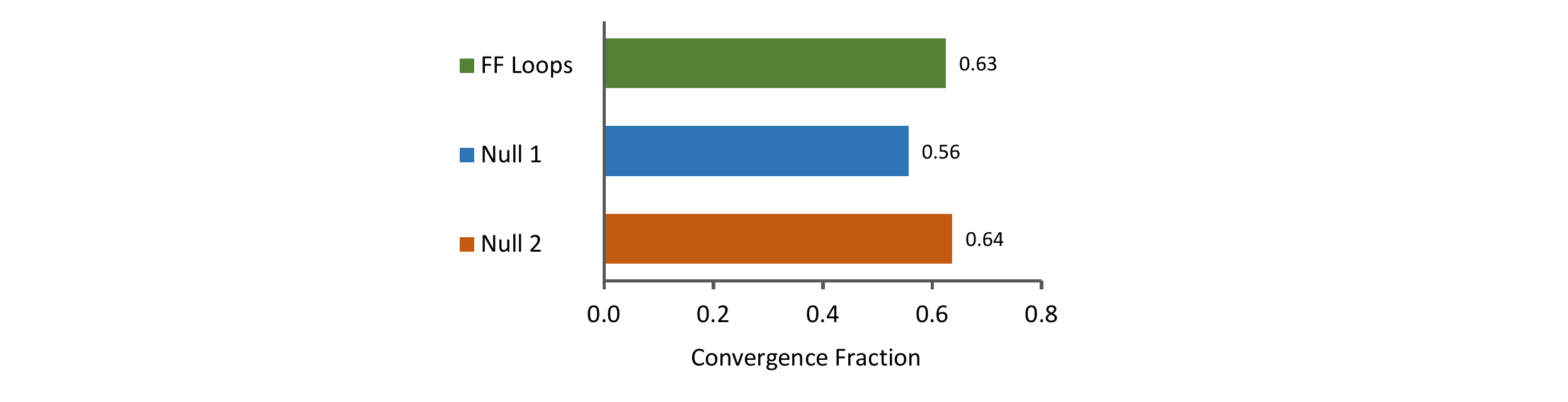}
	\end{center}
	\renewcommand{\baselinestretch}{1}
	\small{
		\textbf{Supplementary Figure \arabic{figure} Convergence of networks enriched with feed-forward loops.}  An ensemble of 500 networks, each enriched with 300 additional feed-forward loops, was tested for convergence (green). Results are compared to random networks with the same degree sequence (blue) and  networks prior to enriching the networks with feed-forward loops (red). Initial networks prior to the addition of FF loops have SF out-degree distributions with $a=1$, $\gamma=2.4$ and Binomial in-degree distribution with  $p= \dfrac{3.5}{N}$ and $N$. Other parameters are $N=1000$, $g_0 = 10$, $\alpha = 100$, $\mathcal M_0 = 2$, $\varepsilon = 3$, $c=0.2$, $D=10^{-3}$ and $y*=0$. }
	
\end{figure}

\par
\bigskip
\noindent
\textbf{\textit{Bi-Fans}} \par
\noindent The four-node motivf which is thought to be most significantly over-represented in regulatory networks is the bi-fan in which two regulators jointly regulate two target genes: $A \Rightarrow C$, $A \Rightarrow D$, $B \Rightarrow C$, $B \Rightarrow D$. We find that over representing this motif does not have strong positive or negative effect compared to the network prior to adding the loops or the null model.

\subsection*{Stretched Exponential Fit to the Distribution of Convergence Times} 
Convergence times of exploratory adaptation can be well fit by a stretched exponential (main text Fig. 4).
The fit is calculated by fitting $1-log(CDF)$ of convergence times to a power law. Thus, the fit to the CDF has the stretched exponential  form $1-e^{-{x/\lambda}^k}$ which implies a Weibull distribution, with PDF

\be
f(t) =
\begin{cases} 
	\hfill \frac{k}{\lambda}{(\frac{t}{\lambda})^{k-1}}e^{-{t/\lambda}^k}    \hfill & x>0 \\
	\hfill 0 \hfill & x\leq0 \\
\end{cases}
\ee

\par For the distributions shown in the main text (Fig. 12), the fit of $1-log(CDF)$ to a power law is excellent with ${R^2}\geq0.995$. To gain further understanding of the distribution of convergence times, we computed the empirical mean and standard deviation for increasingly larger time windows. Results show that both moments monotonically increase with the window size (even for very large time windows - Sup. Fig. 12A and 12B blue lines). In addition we calculated the stretched exponential fit for each time window. We did not use all the data points in each window, but rather a fixed number of 200 data points for all windows which were evenly distributed in the window. Thus we avoid the possibility of erroneous estimations of the quality of the fit stability which might result from an increase in the size of the data set for large time windows. Using this fit protocol we obtained excellent fits of $1-log(CDF)$ to a power law For all time windows above t=3000 (${R^2}\geq0.995$).  We also found that the fit is stable and that after an initial transient the fit parameters  fluctuate very little with increased windows sizes (Sup. Fig. 12A and 12B red lines). These findings increase our confidence in the stretched exponential fit. Moreover, although we observed an increase in mean and std with window size, they do not diverge. The stability of the fit and its large mean and std may indicate that the first two moments of the convergence times distribution are finite but can only be estimated faithfully using much larger time windows.

\begin{figure}[H]
	\refstepcounter{figure}
	\label{fig:eigen}
	\begin{center}
		\includegraphics[width=16cm]{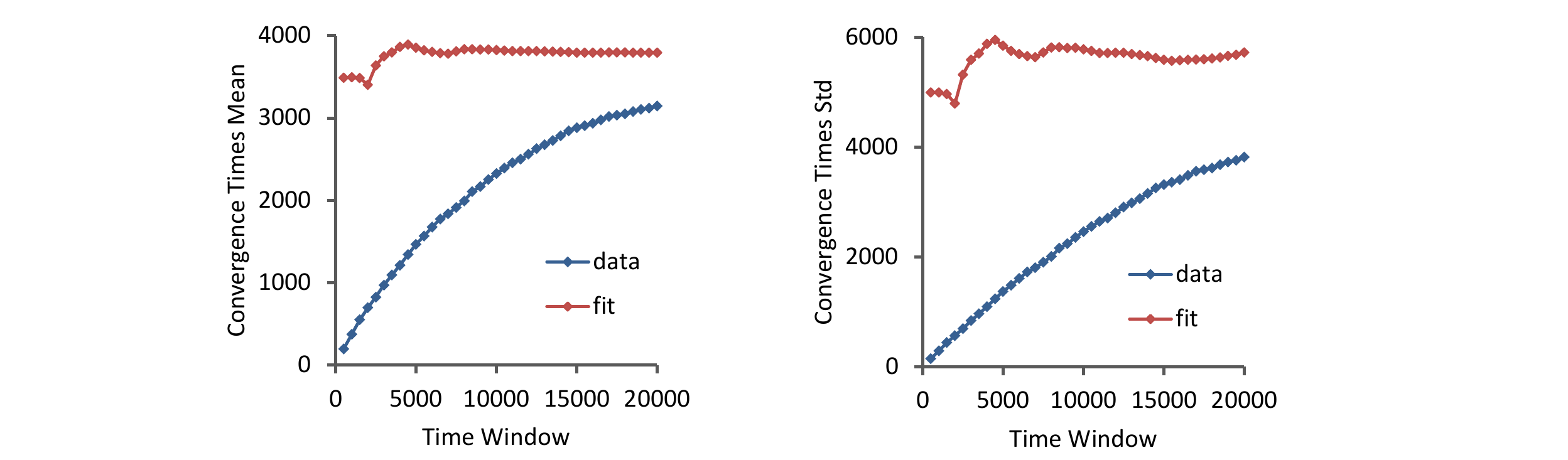}
	\end{center}
	\renewcommand{\baselinestretch}{1}
	\small{
		\textbf{Supplementary Figure \arabic{figure} Stability of stretched exponential fit to the distribution of convergence times.}  the empirical mean and standard deviation for increasingly larger time windows is shown (A and B blue lines). Both monotonically increase with window size for all times tested. Mean and variance estimated from parameters of the stretched exponential fit  (see text for detail) are stable and  after an initial transient fluctuate very little with increased windows sizes (A and B reds). Ensemble has SF out-degree distribution with $a=1$, $\gamma=2.4$ and Binomial in-degree distribution with  $p= \dfrac{3.5}{N}$ and $N$. Other parameters are $N=1500$, $g_0 = 10$, $\alpha = 100$, $\mathcal M_0 = 2$, $\varepsilon = 3$, $c=0.2$, $D=10^{-3}$ and $y*=0$. }
	
\end{figure}  

\subsection*{Stability of the Adapted State}

As shown in the main text, a large fraction of the networks with appropriate topology converge to a stable state for which the phenotype $y$ remains sufficiently close to the demand $y*$. From a biological point of view these final states are likely to be perturbed. Therefore it is of interest to examine the resilience of the final stable state to perturbations both in the nodes' states $x_i$ and the interactions $W_{ij}$. The system at hand is high-dimensional, nonlinear and includes a stochastic feedback in the form of a random-walk in its parameters. These properties make the analytical assessment of the stability of the full model very difficult Therefore, we shall address the question of stability numerically.
\bigskip\\
\noindent
\textbf{\textit{Perturbations to $\mathbf{x}$}} \par
\noindent A standard linear stability analysis can be employed for network parameters $W_{ij}$ at the values they reached following exploration. Examinations of the Jacobian matrix at such a fixed-point reveals that its eigenvalues mostly cluster around -1, with a few outliers eigenvalues (Sup Fig. 13A Inset). This is not surprising given the structure of the equation of motion $\dot{x}_i = -x_i + \sum W_{ij}\phi \left(x_j\right)$. At a fixed-point one may expect the linear term $-x_i$ to dominate typically while the other term may average out, resulting in an eigenvalue which is close to -1. The overall stability of the fixed-point can be quantified by the largest eigenvalue of the Jacobian at the fixed-point. The distribution of the largest eigenvalues, computed across and ensemble of converged networks at their respective fixed-points, are mostly located near -1 as well (Sup Fig. 13A).
While such analysis provides some information as to the stability of fixed-points, it has three major disadvantages: (i) It is limited  to the case where the converged state is a fixed point; (ii) It is relevant only to constant parameters $W_{ij}$, in contrast to the exploratory dynamics we described. Any large enough perturbation in $\mathbf{x}$ will be naturally  accompanied by change in $W_{ij}$ due to the divergence of $y$ from the $\epsilon$ comfort-zone around $y*$.  (iii) Even for constant $W_ij$, linear stability analysis is only valid locally around the fixed-point. For nonlinear dynamics with a large number of dimensions, the basin of attraction around a fixed point may be very small and beyond it  the linear analysis does not hold. In such cases even relatively small perturbation to a system with negative eigenvalues my cause the system to lose its stability. \par 
Given these disadvantages of linear stability analysis, we employ an additional method to assess the stability of the system to a perturbations in $\mathbf{x}$. Stability is assessed numerically by directly employing the full exploratory dynamics to perturbed variables and computing the convergence times. More specifically, we examined an ensemble of 500 networks which converged to a stable state. For each network we perturbed the final state $\mathbf{x}_{converged}$ randomly by 5\%, 10\%, 20\%, 50\%, 100\% and 200\%, and simulated the dynamics with the perturbed state $\mathbf{x}_0 \coloneqq \mathbf{x}_{pertubed}$ and final interaction matrix $W_{converged}$ as initial condition, and the same phenotype vector $\mathbf{b}$.  Convergence times for these simulation were recorded. Results are shown in Sup. Fig. 13B, alongside  a control (labeled "Null") which is composed of an ensemble of 500 networks with the same parameters as the converged networks. 
As can be seen for perturbations of 5\%, nearly all networks re-adapted within 2000 time units and a large number of these networks re-converged very rapidly. For larger perturbation,
we find a higher convergence than in the control and a lager fraction of rapid convergences. This suggests that the stability of the converged state in $\mathbf{x}$ space is non-local and that the basin of attraction covers a large area of the $\mathbf{x}$ space. 
Moreover, the convergence process following the perturbation involves changes of the parameters $W$. The rapid convergence for large number of networks suggests that the phase space of the dynamics in $\mathbf{x}$ deforms continuously with the parameters $W_{ij}$ and that in many cases the existence of a stable attractor in not affected by small perturbations to these parameters.

\begin{figure}[H]
	\refstepcounter{figure}
	\label{fig:eigen}
	\begin{center}
		\includegraphics[width=16cm]{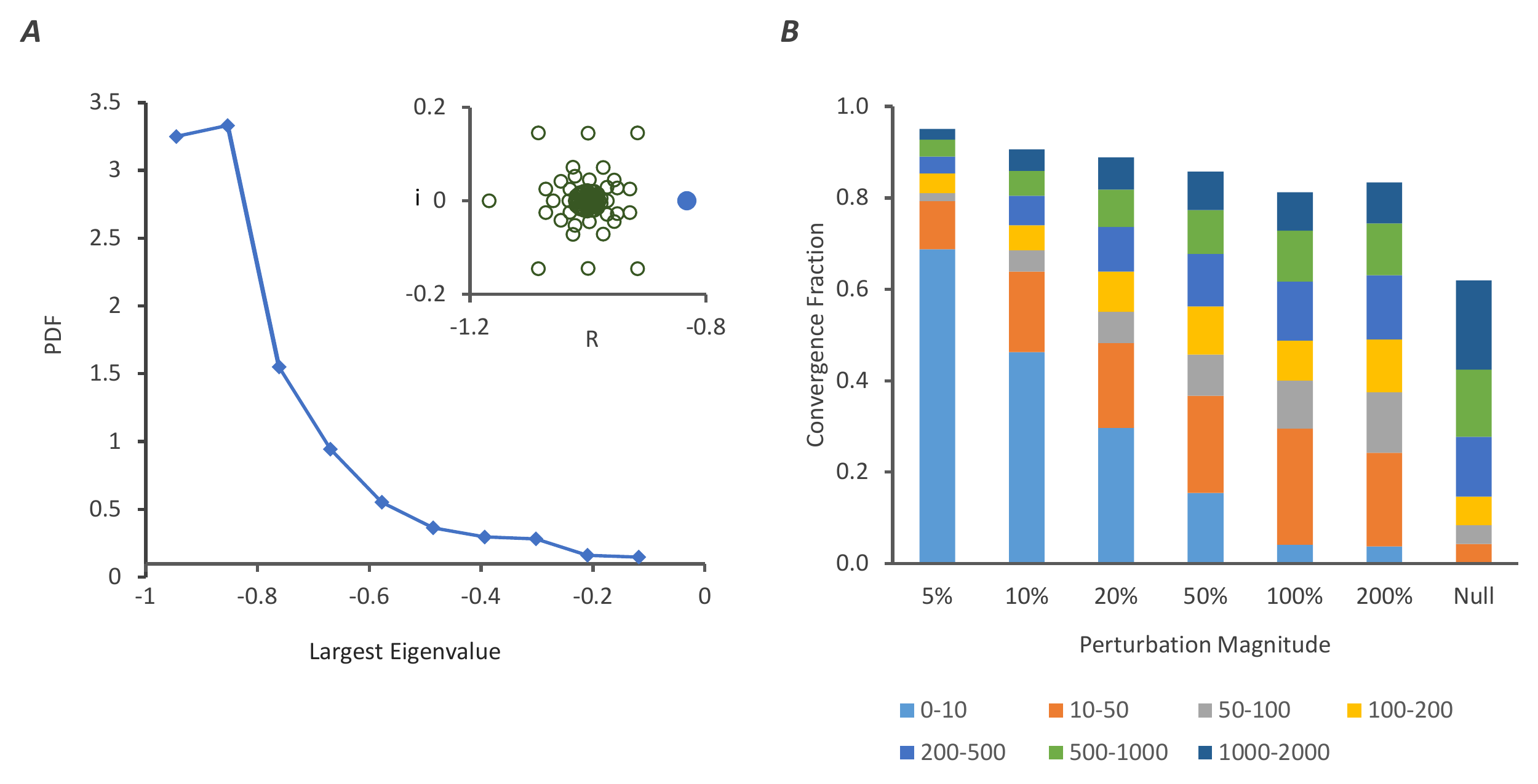}
	\end{center}
	\renewcommand{\baselinestretch}{1}
	\small{
		\textbf{Supplementary Figure \arabic{figure}. Perturbation to the state vector  $\mathbf{x}$ after convergence.}  (A)  PDF of the maximal real-part of eigenvalues of the Jacobian matrix at the fixed-point, across an ensemble of 700 networks. Typical distribution eigenvalues in the complex plane for a single network is shown in the inset. Largest real part of eigenvalue in blue.  (B) Convergence fraction of networks for which the final state $\mathbf{x}_{converged}$ was perturbed and a null control.  All networks have SF out-degree distributions with $a=1$, $\gamma=2.4$ and Binomial in-degree distribution with  $p= {3.5}/{N}$ and $N$. Other parameters are $N=1000$, $g_0 = 10$, $\alpha = 100$, $\mathcal M_0 = 2$, $\varepsilon = 3$, $c=0.2$, $D=10^{-3}$ and $y*=0$. }
	
\end{figure}

\noindent \textbf{\textit{Perturbations to $W_{ij}$}} \par
\noindent In order to asset the effect of perturbations to $W_{ij}$ we examined 10 distinct networks $\left\lbrace W^1,..W^{10}\right\rbrace $ with SF-Binom connectivity after they converged to an adapted state. For each network $W^i$ we perturbed all non zero connections $W_{ij}$ randomly by 1\% 250 times thus obtaining 250 new networks. Simulations were then run for these 250 perturbed networks using the same phenotype vector $\mathbf{b}$ used for $W^i$ and initial conditions $\mathbf{x}_0$ equal to the converged state $\mathbf{x}_{converged}$ of the network $W^i$. Re-adaptation convergence times for these perturbed networks were then recorded. This protocol was repeated for perturbations of magnitude 5\%, 10\%, 20\% and 50\%, using as a basis the same converged network $W^i$. In addition we constructed a null control for each converged network $W^i$ by running 500 simulation with random choices of $W_{ij}$ and $x_i$, while using the same backbone $T^i$ which corresponds to $W^i$. Thus the statistics of convergence for the perturbed networks  can be compared to random networks with the same backbone $T^i$. All of these simulation were repeated for each of the 10 networks $\left\lbrace W^1,..W^{10}\right\rbrace $. The results, averaged over the 10 networks, are shown in Sup Fig. 14.
As can be seen in Sup Fig. 14A, for small perturbations of 1\% nearly all networks re-adapted within 2000 time units and a large number of these networks re-converged very rapidly. For larger perturbation
(5\% and 10\%) convergence was at lower fractions and less rapid but still more than the control (Sup. Fig. 14A.). These findings suggest a continuous picture: small perturbations re-converge rapidly in high fractions, intermediate perturbations (20\%) less so, and for large perturbations (50\%) convergence statistics is similar to that of random networks with the same backbone. Consistently with this picture, Sup. Fig. 14B shows that the coordinates of the new fixed-points move away from the original one in a continuous manner.

\begin{figure}[H]
	\refstepcounter{figure}
	\label{fig:eigen}
	\begin{center}
		\includegraphics[width=16cm]{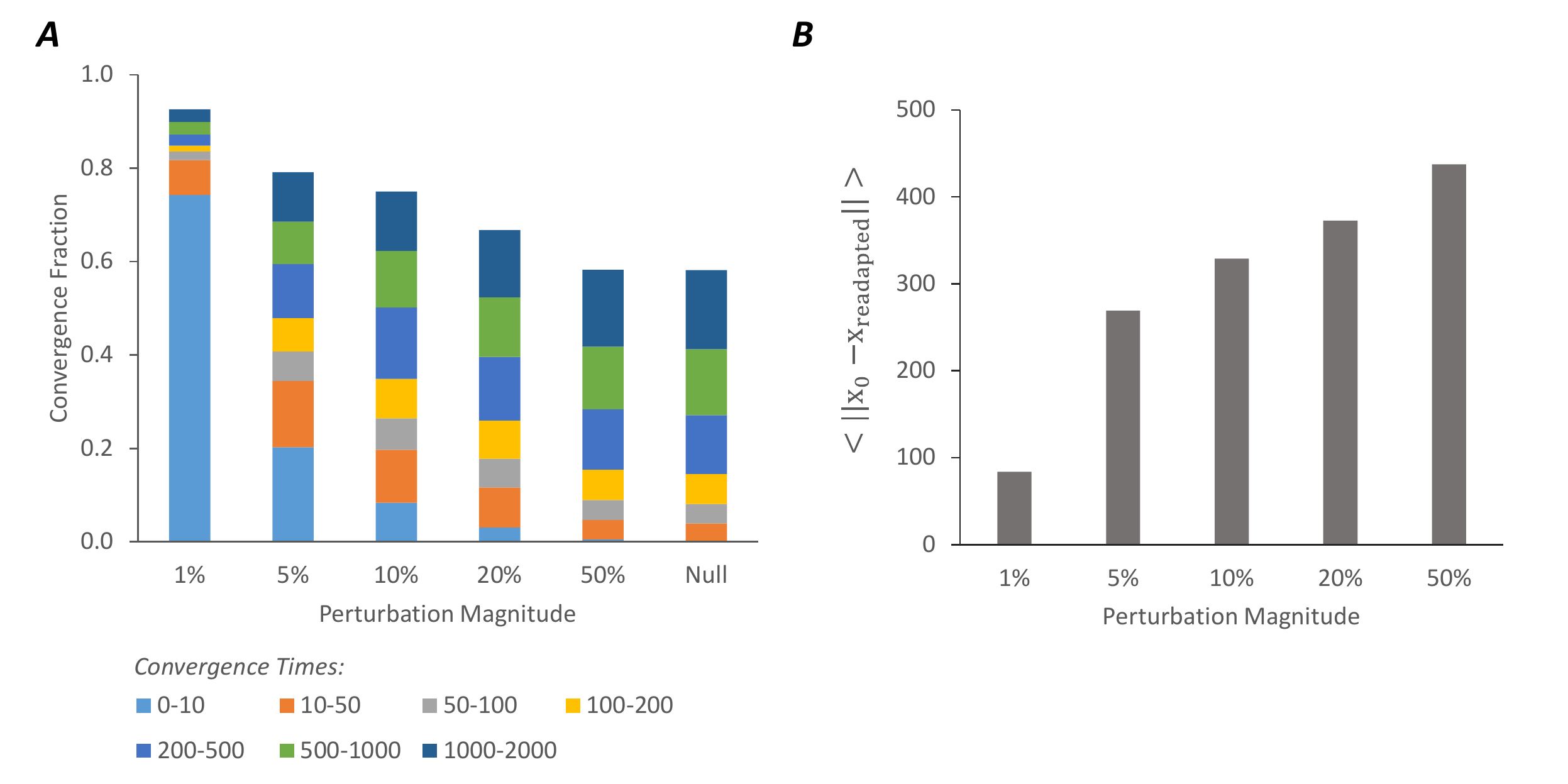}
	\end{center}
	\renewcommand{\baselinestretch}{1}
	\small{
		\textbf{Supplementary Figure \arabic{figure}. Perturbation to network connection strengths after convergence.}  Non-zero connection strengths, $W_{ij}$, of 10 networks were randomly perturbed following convergence to fixed points. For each perturbation size an ensemble of 250 networks was constructed. In addition, a control ensemble was constructed consisting of 500 networks with random connection strengths $W_{ij}$ and the same topological backbone as the original network. Results are averaged over the 10 networks.(A)  Convergence fraction of perturbed networks and null control. (B) Average Euclidean distance between the initial converged state $\mathbf{x}_0$ before perturbation and the converged state after the perturbation $\mathbf{x}_{readapted}$. All networks have SF out-degree distributions with $a=1$, $\gamma=2.4$ and Binomial in-degree distribution with  $p={3.5}/{N}$ and $N$. Other parameters are $N=1000$, $g_0 = 10$, $\alpha = 100$, $\mathcal M_0 = 2$, $\varepsilon = 3$, $c=0.2$, $D=10^{-3}$ and $y*=0$. }
	
\end{figure} 
\noindent
\textbf{\textit{Perturbations to $T$}} \par
\noindent The effect of perturbations to the backbone $T$ were assessed similarly manner to the perturbations in $W_{ij}$ described above. We examined 10 distinct networks $\left\lbrace W^1,..W^{10}\right\rbrace $ with SF-Binom connectivity after convergence; each backbone $T$ was perturbed by adding or deleting a random fraction of connections to the network. Each such perturbation was applied 250 times and simulations were then run with the same phenotype vector $\mathbf{b}$ used for $W^i$. Initial conditions $\mathbf{x}_0$ for these runs were the converged state reached for $W^i$ prior the perturbation. For new connections $T_{ij} = 1$ that were added, the statistics of the connection strength $W_{ij}$ was chosen randomly from the same distribution as the existing connection strengths in the network.  \par
The results, averaged over the 10 networks, are shown in Sup Fig. 15. 
For both additions and deletions convergence is relatively stable for small perturbations, And for such perturbations many network quickly return to a converged state (Sup Fig. 15A and 15B). 

\begin{figure}[H]
	\refstepcounter{figure}
	\label{fig:eigen}
	\begin{center}
		\includegraphics[width=16cm]{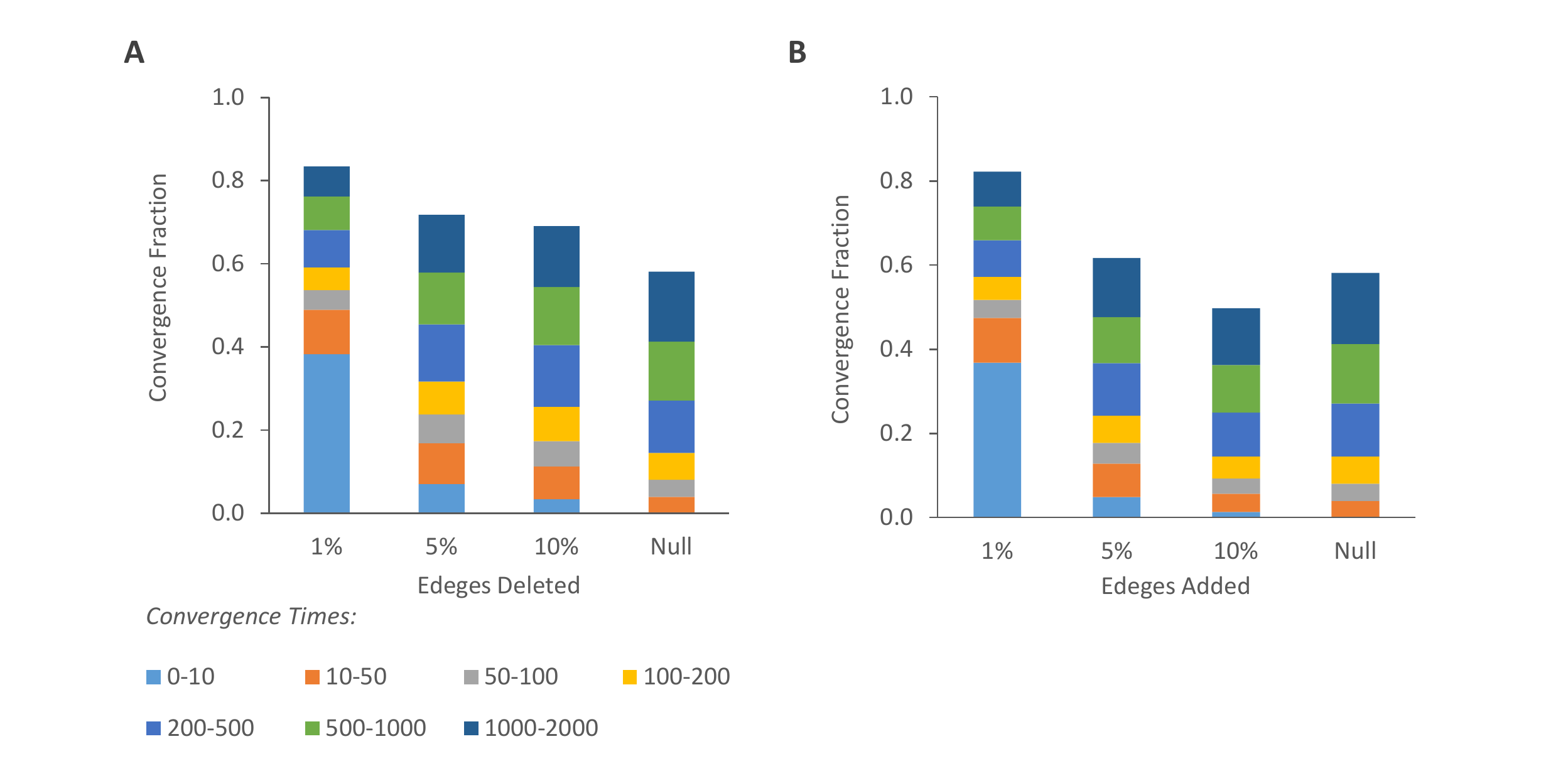}
	\end{center}
	\renewcommand{\baselinestretch}{1}
	\small{
		\textbf{Supplementary Figure \arabic{figure} Perturbation to topology of converged networks.}  Topology, $T$, of 10 converged networks was randomly perturbed by deleting and adding connections. A varying number of random connections have  been removed (A) or added (B) to the converged networks. For each magnitude of perturbation an ensemble of 250 perturbed networks was constructed. Network in these ensembles were then simulated and convergence times were tracked. Results are averaged over the 10 initial networks. Initial converged networks have SF out-degree distributions with $a=1$, $\gamma=2.4$ and Binomial in-degree distribution with  $p= {3.5}/{N}$ and $N$. Other parameters are $N=1000$, $g_0 = 10$, $\alpha = 100$, $\mathcal M_0 = 2$, $\varepsilon = 3$, $c=0.2$, $D=10^{-3}$ and $y*=0$. }
	
\end{figure}

\newpage
\section*{\centering \huge {Supplementary Note 4}  }
\vspace{10ex}
\setcounter{equation}{0}

\subsection*{Dependence of Convergence of Fixed Networks on Largest Hub}
We have seen that convergence of exploratory adaptation correlates with the fraction of fixed networks (constant networks and no constraint - Fig. 7 in the main text) in which the intrinsic dynamics of Eq. (1) converges to fixed points. Here we investigate further the dependence of constant networks   on the network hubs. In particular, we ask whether the existence of larger hubs in a network correlates with larger probability of convergence to fixed point. In order to examine this property we randomly constructed  backbones $T$ of size N=1500 with SF out-degree and Binomial in-degree distributions. We picked 20 such backbones $\{T_1...T_{20}\}$ for which the largest hub has outgoing degrees between ${{K_1}} \sim 100$ and  ${{K_{20}}} \sim 1100$. Next we created for each backbone $T_i$ an ensembles of 500 networks, each with random interactions strengths $\{J^{i,1}...J^{i,500}\}$, $1\leq i \leq 20$. For each ensemble we computed the fraction of networks which converged to a fixed-point in the open-loop setting.  Sup. Fig. 16 depicts this fraction as a function of the largest degree, showing a noisy but significant correlation. The large fluctuation indicate that there are other properties in addition to the largest hub that have a significant influence on convergence. 
\begin{figure}[H]
	\refstepcounter{figure}
	\label{fig:eigen}
	\begin{center}
		\includegraphics[width=13cm]{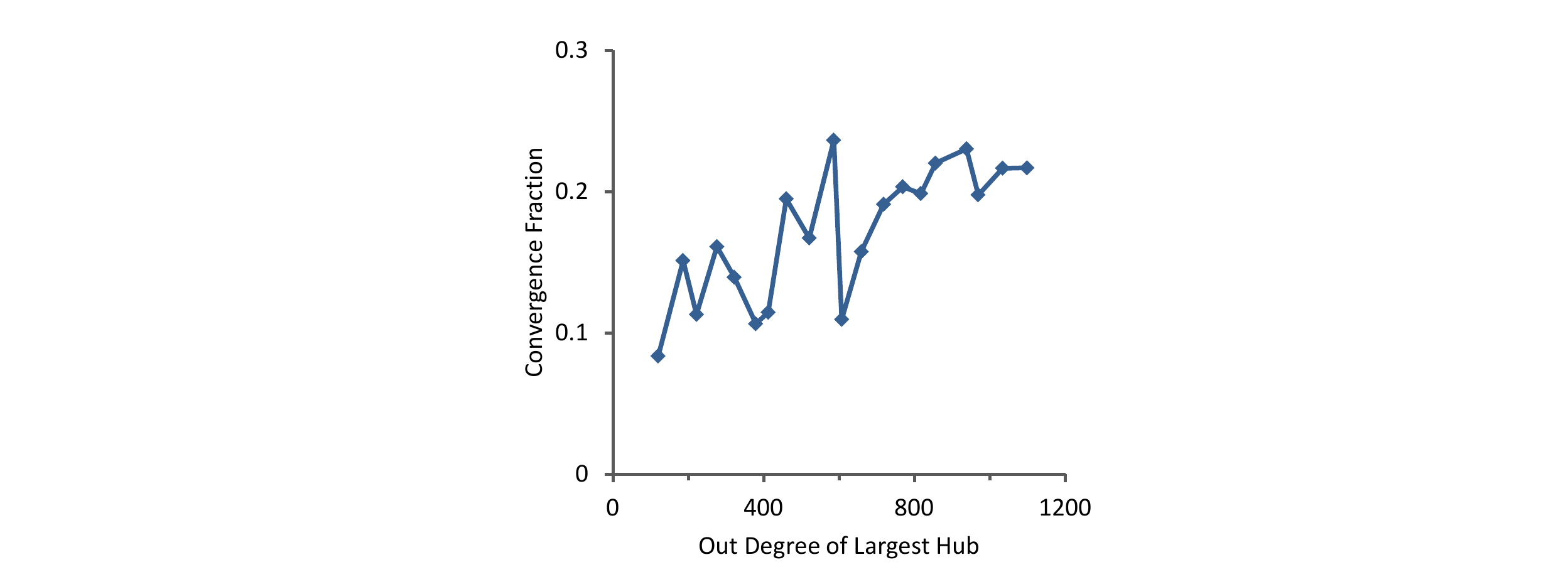}
	\end{center}
	\renewcommand{\baselinestretch}{1}
	\small{
		\textbf{Supplementary Figure \arabic{figure}. Dependence of convergence fractions to fixed points in constant networks on largest hub.}  Twenty backbones $\{T_1...T_{20}\}$ were used to generate 20 ensembles, each composed of 500 random realizations of interaction strengths $J$. Each backbone $T_i$ has a different maximal degree of the largest out-going hub, between ${{k_1}} \sim 100$ and  ${{k_{20}}} \sim 1100$. Each data point represents the fraction of networks within the ensemble which converged to a fixed-point within a time window of 2000 in the ensemble, plotted as a function of the maximal out-degree. All backbones are drawn from a SF out-degree distribution with  $a=1$, $\gamma=2.4$ and Binomial in-degree distribution with  $p\simeq \dfrac{3.5}{N}$ and $N$ . $N=1500$ and $g_0 = 10$.}
\end{figure}

\subsection*{Dependence of Convergence of Fixed Networks on Network Gain}
Convergence fractions under exploratory adaptation dynamics are weakly dependent on the network gain $g$ (Fig. 2 D in the main text). We examined the analogous property for constant networks by randomly constructing 5 backbones with SF out-degree and Binomial in-degree, $\{T^1...T^{5}\}$. For each backbone $T^i$ we created seven ensembles of 500 networks, each with a different g, $\{(T^i,J_g^j)\}_{j=1}^{500}$, $1\leq i \leq 5$ , $g\in\{2,3,6,8,10,12,15\}$ (a total of 35 ensembles). For each ensemble we computed the fraction of networks for which the intrinsic dynamics converges to a fixed-point (fixed network no constraint - Sup Fig. 17 doted lines). In addition we averaged over backbones by constructing seven ensembles with $g\in\{2,3,6,8,10,12,15\}$ in which each network has a different $T$ and $J$,  $\{(T^j,J_g^j)\}_{j=1}^{500}$ (Sup. Fig. 17 dark blue). Both types of ensembles show a weak dependence on $g$ after an initial decline for small $g$ . In addition convergence fractions are also dependent on the specific topology $T^i$.     

\begin{figure}[H]
	\refstepcounter{figure}
	\label{fig:eigen}
	\begin{center}
		\includegraphics[width=13cm]{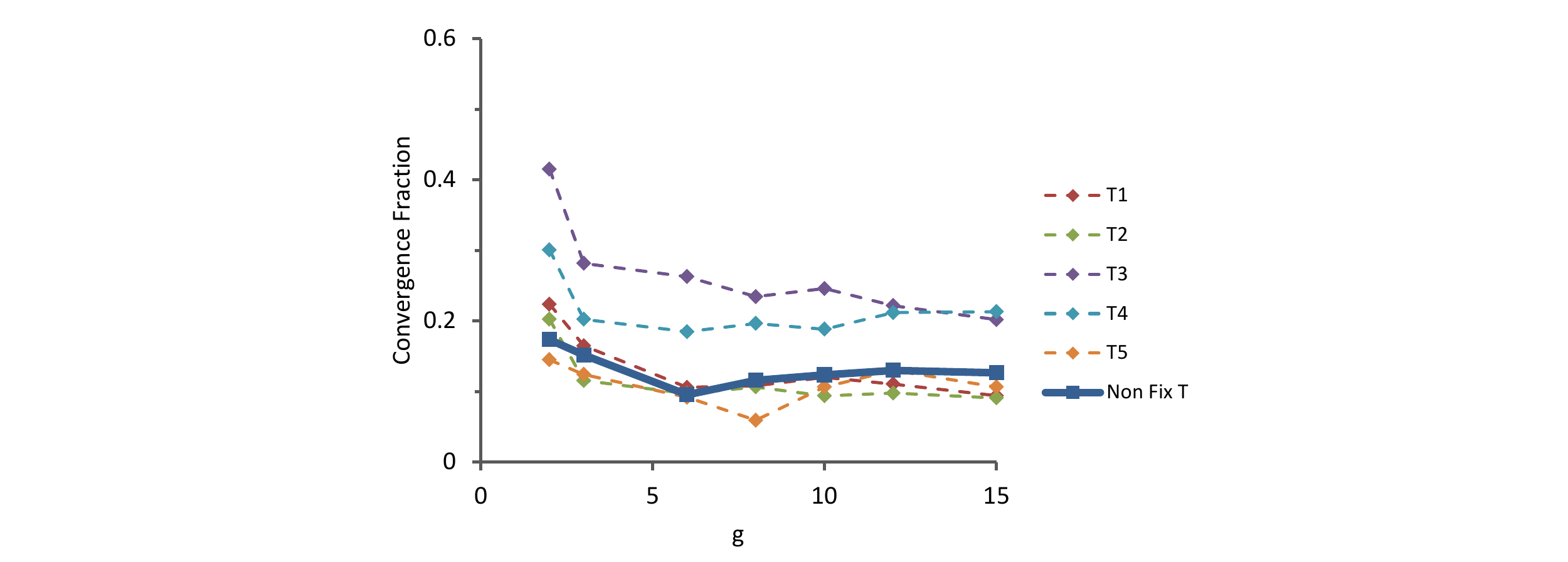}
	\end{center}
	\renewcommand{\baselinestretch}{1}
	\small{
		\textbf{Supplementary Figure \arabic{figure}. Dependence of convergence fractions to fixed points in constant networks on network gain.}  Ensembles of 500 networks with fixed $T^i$, and  different g, $\{(T^i,J_g^j)\}_{j=1}^{500}$, $1\leq i \leq 5$ , $g\in\{2,3,6,8,10,12,15\}$ (doted lines), were tested for converged to a fixed-point with a fixed network and no constraint. Mixing of the different backbones into ensembles characterized by $g$ results in the thick blue line. Ensembles have SF out-degree distribution with  $a=1$, $\gamma=2.4$ and Binomial in-degree distribution with  $p\simeq \dfrac{3.5}{N}$ and $N$, $N=1500$.
	}
\end{figure}

\bibliographystyle{unsrt}
\bibliography{suppbib} 


\end{document}